\documentstyle[prc,epsf,aps]{revtex}

\begin{document}
\title{Meson and Isobar Degrees of Freedom in A($\vec{e},e'\vec{p}$)
reactions at $0.2 \leq Q^2 \leq 0.8$~(GeV/c)$^2$} 
\author{Jan Ryckebusch \footnote{E-mail :
jan.ryckebusch@rug.ac.be},  Dimitri Debruyne, Wim Van Nespen and Stijn
Janssen}
\address{Department of Subatomic and Radiation Physics \protect\\
University of Gent, Proeftuinstraat 86, B-9000 Gent, Belgium}

\date{\today}
\maketitle
\begin{abstract}
The effect of meson and isobar degrees of freedom in
A($\vec{e},e'\vec{p}$) and A($e,e'n$) is studied for four-momentum
transfers $Q^2$ in the range between 0.2 and 0.8~(GeV/c)$^2$.  The
calculations are performed in a non-relativistic framework with
explicit $(N,\Delta,\pi)$ degrees-of-freedom.  For the whole range of
momentum transfers under investigation the relative effect of the
meson-exchange and isobar degrees of freedom is significant.  At low
missing momenta and quasi-elastic conditions, a tendency to reduce the
$(e,e'p)$ and $(e,e'n)$ differential cross sections is noticed.  The
greatest sensitivity is found in the interference structure functions
$W_{LT}$ and $W_{TT}$.  The recoil polarization observables, on the
other hand, are moderately affected by the meson-exchange and
$\Delta$-isobar currents.

\end{abstract}


\section{Introduction}

Systematic investigations of A($e,e'p$) reactions at Saclay, NIKHEF,
Mainz and Bates have produced an impressive amount of data for various
target nuclei \cite{kelly96}.  For the sake of minimizing the
uncertainties with respect to the reaction mechanism, a large fraction
of these investigations were done under or close to quasi-elastic
kinematics (Bjorken $x= \frac {-q ^\mu q _ \mu } {2M_N \omega }
\approx$~1).  These data sets highlight at the same time the success
and the limits of the independent-particle-model (IPM) for atomic
nuclei \cite{vijay98}.  Indeed, after corrections for final-state
interactions and Coulomb distortions of the electron probe, the shape
of the deduced proton momentum distributions are systematically in
line with the predictions of modern formulations of the nuclear IPM.
On the other hand, below the Fermi momentum the absolute magnitude of
the deduced momentum distributions are systematically,
i.e. independent of the nucleon momentum, lower than IPM
predictions. To cut a long story short, the major conclusion from this
world-wide $(e,e'p)$ effort seems that an appropriate non-relativistic
picture of the
nucleus is roughly compatible with {\sl 70\% mean-field behaviour and 30\%
``correlations''} an observation which is still frequently ignored in
various nuclear structure calculations and model developments.  The
energy of the available electron beams with a large duty factor
($\epsilon \leq $~1~GeV) made that most of this aforementioned
$(e,e'p)$ work was done at four-momentum transfers of the order $Q^2=
-q ^\mu q _ \mu \leq$~0.2~(GeV/c)$^2$. With the advent of the TJNAF
and an upgraded Mainz electron facility higher values of $Q^2$ come
into reach of experimental exploration.

Amongst the major physics' goals motivating exclusive ($e,e'p$)
measurements from finite nuclei at higher momentum transfer ($Q^2 \geq
0.2~$ (GeV/c)$^2$) one can mention the following ones. The higher
$Q^2$ conditions and unmistakingly smaller distance scales probed
should make it feasible to achieve a better understanding of the
short-range mechanisms in nuclei.  At the same time, one could hope to
find experimental evidence for the onset of quark and gluon degrees of
freedom.  Presumably, the most convincing evidence pointing into that
direction could come from measurements for processes that are fairly
well understood at lower $Q^2$ (like ($e,e'p$) at x=1) and turn out to
be completely at odds with meson/baryon models when higher $Q^2$
regimes are entered.  Another challenge for ($e,e'p$) measurements at
higher momentum transfers is the question whether available
relativistic models can succeed in producing a better agreement with
the data sets than the non-relativistic ones and if so, which
dynamical degrees of freedom make them to be substantially different
from what is commonly implemented in the non-relativistic nuclear
many-body models. Another fundamental question that has received a
great share of attention for many years, is the question whether the
nucleon properties (like electromagnetic form factors e.g.) are
modified in the nuclear medium.  This information is of unvaluable
importance for models that embark on the ambitious program of
understanding nuclei in terms of quark and gluon degrees of freedom
\cite{lu98a,lu98b}.  A challenging but at the same time rather
ambiguity-free way of probing the medium-dependent form factors, are
double polarization observables from A($\vec{e},e'\vec{p}$)
measurements \cite{laget94,e89033,e93049,e91006}.  Indeed, double
polarization observables are conceived to be rather insensitive to
ambiguities with respect to the final state interactions (FSI) that
affect most of the other A($e,e'p$) observables.  Other effects that
are recognized to possibly complicate the interpretation of double
polarization observables in terms of medium-dependent nucleon form
factors are gauge ambiguities, channel-couplings and two-body current
effects.  Whereas the first two sources of possible dilutions were
extensively studied by Kelly \cite{kelly1,kelly2} and found to produce
very small corrections, the two-body current effects are not well
studied for finite nuclei.

Earlier efforts to study the role of two-nucleon currents in exclusive
A($e,e'p$) reactions from finite nuclei (A$\geq$4) include the
pioneering work of Suzuki \cite{suzuki89}, the systematic
investigations by the Pavia group \cite{boffi90,boffi90a,boffi92} and
the $^4$He($e,e'p$) studies as e.g. reported in
Refs.~\cite{epstein93,laget94}.  In order to make their calculations
computationally more attractive, the results of the Pavia group were
obtained with a two-body current operator that was formally reduced to
an effective one-body one.  This approximation adopts a Fermi-gas
picture for the residual nucleus that allows to integrate out the
coordinates of the second nucleon that gets involved in the
photoabsorption process.  For the work presented here, we treat the
two-body currents in their full (non-relativistic) complexity and deal
with the multi-dimensional integrals that automatically occur when
several nucleons get involved in the (virtual) photoabsorption
process.  A similar sort of exact treatment was earlier adopted in our
work reported in References \cite{veerle1,veerle2}.  There it was
found that the effect of the two-body currents on the A($e,e'p$) cross
sections, just as the role of the coupled-channel effects, is
gradually decreasing with increasing momentum transfer. This conforms
to the findings with respect to the momentum-transfer dependence of
the two-body current effects in the d($e,e'p$)n and $^4$He($e,e'p$)
\cite{laget94}.  Nevertheless, even at the highest momentum transfer
considered in Ref.~\cite{veerle1} (q=600~MeV/c) the cross sections
were predicted to be substantially affected by the two-body currents.
In this work the role of meson-exchange and isobar degrees of freedom
in A($\vec{e},e'\vec{p}$) and A($e,e'n$) is investigated in a wide
range of four-momentum transfers (0.2 $\le Q^2 \le$0.8~(GeV/c)$^2$).
Special emphasis is placed on the recoil polarization observables as
they open perspectives to investigate possible medium modifications of
the nucleon properties.

The calculational framework will be introduced in
Section~\ref{sec:model}.  In Section~\ref{sec:observ} the adopted
conventions for the ($\vec{e},e'\vec{p}$) observables will be
introduced. The model for the bound and scattering states is described
in Section~\ref{sec:fsi}.  In Section~\ref{sec:current} the model
assumptions with respect to the one- and two-body current operators
will be summarized. The results of the calculations are contained in
Section~\ref{sec:results}. Our conclusions are summarized in
Section~\ref{sec:conclusions}.

\section{Theoretical framework}
\label{sec:model}
\subsection{A($\vec{\mathrm{e}}$,e$'\vec{\mathrm{p}}$) observables and
kinematics}
\label{sec:observ}

\begin{figure}
\centerline{\epsfysize=10.cm \epsfbox{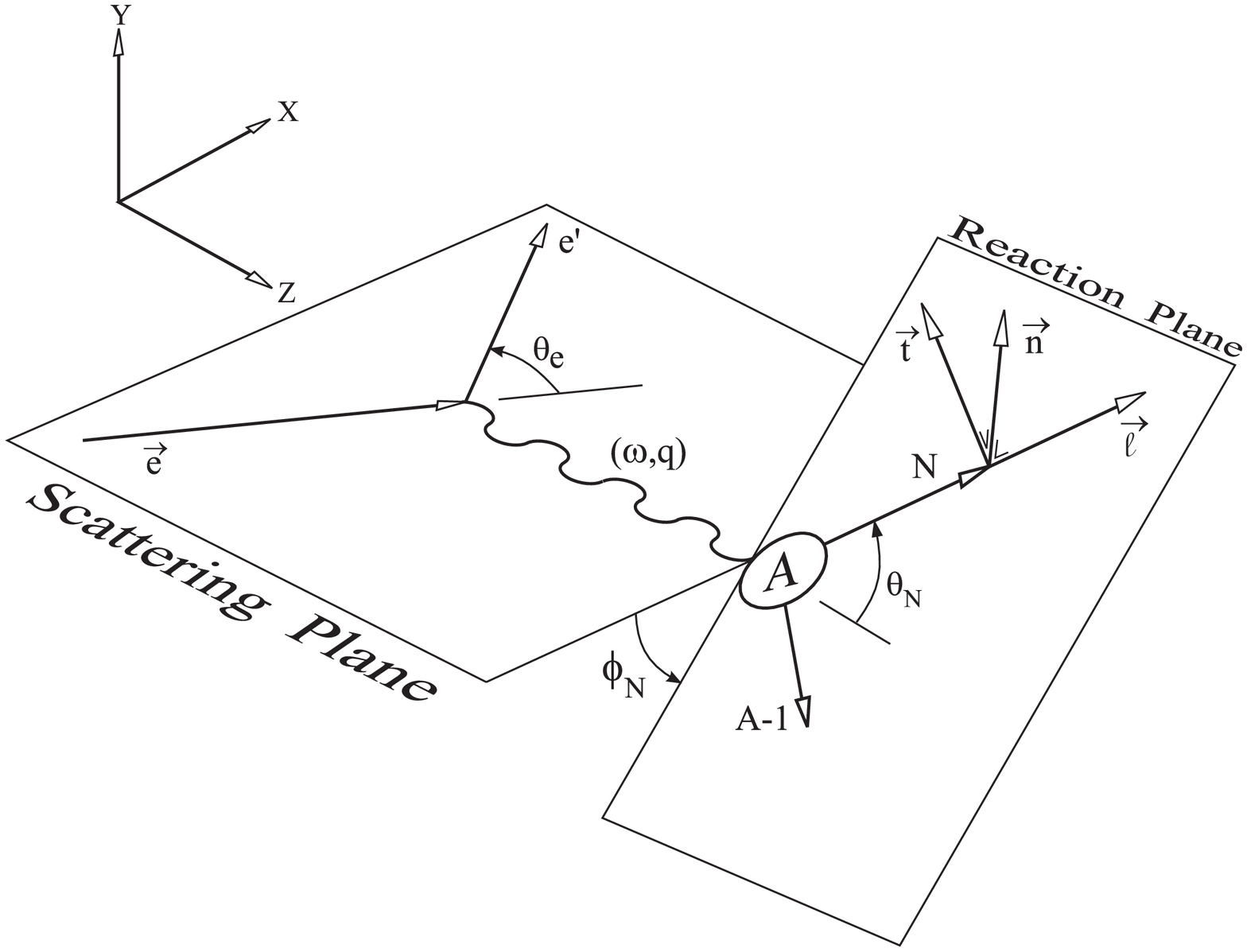}}
\caption{Diagram of the ($e,e'p$) process showing the electron
scattering plane, the reaction plane and the basis in which the
ejectile's polarization is determined.}
\label{fig:reffram}
\end{figure}

We consider processes in which a longitudinally polarized
electron impinges on a nucleus and induces the following reaction  
\begin{equation}
A \; + \vec{e} (\epsilon) \longrightarrow (A-1)(E_{A-1},\vec{p}_{A-1}) \;
+ N(E_N,\vec{p}_N) \; + e (\epsilon ') \; ,
\end{equation}
to occur.  For such a process the cross section reads in the one
photon exchange approximation
\begin{eqnarray}
& & {d^5 \sigma \over  d \Omega _N d \epsilon ' d \Omega
_{\epsilon '}}  (\overrightarrow{e},e'N)  =  
{1 \over 4 (2\pi)^5 } p_N  E_N  f_{rec}^{-1} \sigma_{M}
\nonumber \\ 
\times & & \Biggl[ v_T {W_T}
+ v_L    {W_L}
+ v_{LT} {W_{LT}}
+ v_{TT} {W_{TT}} 
 + h \biggl[ v'_{LT} {W'_{LT}} + 
v'_{TT} {W'_{TT}} \biggr] \Biggr] \; ,
\label{eq:eep}
\end{eqnarray}
where $f_{rec}$ is the recoil factor 
\begin{equation}
f_{rec} = \left| 1 + \frac {E_N} {E_{A-1}} \left(1 - 
 \frac {\vec{q} \cdot \vec{p}_N} {p_N^2} \right) \right| \; ,
\end{equation}
and $\sigma _M$ the Mott cross section
\begin{equation}
\sigma _M = \frac {\alpha^2} {4 \epsilon ^2} 
\frac {cos ^2 \frac {\theta _e} {2}} {sin ^4 \frac {\theta _e} {2}} \: . 
\end{equation}
The electron kinematics is contained in the kinematical factors
\begin{eqnarray}
v_T &=& tg^2 \frac{\theta_e}{2} -
\frac{1}{2}\left(\frac{q_{\mu}q^{\mu}}{\vec{q}^2}\right)  \\
v_L &=& \left( \frac{q_{\mu}}{\vec{q}} \right)^4
\\ 
v_{LT}&=&\frac{q_{\mu}q^{\mu}}
{\sqrt{2} \mid \vec{q} \mid ^3} (\epsilon + \epsilon ')
tg \frac{\theta_e}{2}
\\ 
v_{TT}&=&\frac{q_{\mu}q^{\mu}}{2 \vec{q} ^2} 
\\
v'_{LT} & = & \frac {q_{\mu}q^{\mu}} {\sqrt{2} \vec{q}^2}  
tg \frac{\theta_e}{2} 
\\
v'_{TT} & = & \sqrt{\left( -\frac {q_{\mu}q^{\mu}} {\vec{q}^2}  +
tg ^2 \frac{\theta_e}{2} \right)} tg \frac {\theta_e} {2}
\; ,
\end{eqnarray}
whereas the structure functions are defined in the standard fashion
\begin{equation}
\begin{array}{ll}
W_L = \left( J_0^{fi} \right)^* \left( J_0^{fi} \right) &
W_T = \left( J_{+1}^{fi} \right)^* \left( J_{+1}^{fi} \right) + 
      \left( J_{-1}^{fi} \right)^* \left( J_{-1}^{fi} \right) \\
W_{LT}= 2 {\cal R}e \left[ \left( J_{0}^{fi} \right)^* \left( J_{-1}^{fi} \right) - 
      \left( J_{0}^{fi} \right)^* \left( J_{+1}^{fi} \right) \right] &
W_{TT}= 2 {\cal R}e \left[ \left( J_{-1}^{fi} \right)^* \left
( J_{+1}^{fi} \right) \right] \\
W_{LT}^{'} = -2 {\cal R}e \left[ \left( J_{0}^{fi} \right)^* \left( J_{+1}^{fi} \right) + 
      \left( J_{0}^{fi} \right)^* \left( J_{-1}^{fi} \right) \right] &
W_{TT}^{'} = \left( J_{+1}^{fi} \right)^* \left( J_{+1}^{fi} \right) - 
      \left( J_{-1}^{fi} \right)^* \left( J_{-1}^{fi} \right) \; . \\  
\end{array} 
\end{equation}
The above definitions for the structure functions and the kinematical
variables correspond with those of Ref.~\cite{raskin}.  We remind that
apart from a negligible parity-violating component, the structure
function W$'_{TT}$ vanishes identically and the 
W$'_{LT}$ in coplanar kinematics if no recoil polarization is
determined.  A new set of
observables comes into reach of experimental exploration when
performing polarimetry on the ejected hadron.  This results in
knowledge about the spin orientation of the ejectile and as
e.g. illustrated in neutron form factor studies represents a powerful
tool to address fundamental physical quantities.  The formal framework
for the electroproduction of polarized nucleons from nuclei is
outlined in great detail in Refs.~\cite{raskin,pickle87,boffigiusti}
and will not be repeated here.  Here, we only review some basic
concepts which mainly serves at introducing the conventions adopted.
For the results presented below, the polarization of the escaping
nucleon is expressed in the so-called barycentric reference frame that
is defined by the following set of unit vectors (Fig.~\ref{fig:reffram})
\begin{eqnarray}
\hat{\vec{l}} & = & \frac {\vec{p}_N} {\left| \vec{p}_N \right|}
\\ \nonumber
\hat{\vec{n}} & = & \frac {\vec{q} \times \vec{p}_N} 
{\left| \vec{q} \times \vec{p}_N \right|}
\\ \nonumber
\hat{\vec{t}} & = & \hat{\vec{n}} \times \hat{\vec{l}} \; .
\end{eqnarray} 
Note that for coplanar kinematics $\hat{\vec{n}}$ determines the y
axis of the reference frame.
The escaping nucleon polarization observables can be determined through
measuring {\bf ratios}.  The {\sl induced polarization} can be addressed
with unpolarized electrons (i=n,l,t) 
\begin{equation}    
{\mathrm P}_i =   \frac {\sigma (s_{N}^{i}= \uparrow) -  
\sigma (s_{N}^{i}= \downarrow)} 
{\sigma (s_{N}^{i}= \uparrow) + \sigma (s_{N}^{i}= \downarrow)}
\hspace{0.5cm} \; ,
\label{eq:indu}
\end{equation}
whereas the {\sl polarization transfer} also requires polarized electron
 beams (i=n,l,t)
\begin{eqnarray}
 {\mathrm P}_i' & = & \frac {
\left[ \sigma (h=1,s_{N}^{i}= \uparrow) - \sigma (h=-1,s_{N}^{i}= \uparrow) \right] 
- \left[ \sigma (h=1,s_{N}^{i}= \downarrow) - 
\sigma (h=-1,s_{N}^{i}= \downarrow) \right] }
{
\left[ \sigma (h=1,s_{N}^{i}= \uparrow) + \sigma (h=-1,s_{N}^{i}= \uparrow) \right] 
+ \left[ \sigma (h=1,s_{N}^{i}= \downarrow) +  
\sigma (h=-1,s_{N}^{i}= \downarrow) \right] }  \; ,
\label{eq:trans}
\end{eqnarray}
where {$s_{N}^{i}= \uparrow$} denotes that the ejected hadron is
spin-polarized in the positive i direction (i=(n,l,t)) and $h$ is the
helicity of the electron impinging on the target nucleus. $\sigma
(h,s_{N}^{i})$ is a shorthand notation for the differential cross
section for an electrodisintegration process initiated by an electron
with helicity $h$ and for which the ejectile is detected with a spin
polarization characterized by $s_{N}^{i}$.  Throughout this work we
adopt relativistic kinematics.  

\subsection{Final-state interactions} 
\label{sec:fsi}

For the model calculations presented here, the bound and scattering
states are produced by solving a Schr\"{o}dinger equation with a
mean-field potential that is obtained from a Hartree-Fock calculation.
The latter are performed with an effective nucleon-nucleon force of
the Skyrme type.  The mean-field potential includes central,
spin-orbit and Coulomb terms.  This model does not require any
empirical input with respect to the inital and final-state potentials.
Moreover, the orthogonality condition between the initial and final
states is obeyed and gauge invariance is preserved at the one-body
current level.  A drawback of the model is that the inelastic
processes, which are commonly accounted for through imaginary parts in
the final-state potential, are only partially included.  The major
impact of the rescattering processes in exclusive ($e,e'p$) is
generally conceived, however, to be a mere reduction of the absolute
cross section.  In the kinematics regime of interest here, the
reduction factor can be estimated from nuclear transparency $(e,e'p)$
measurements as they were recently performed at TJNAF for various
target nuclei \cite{potterfeld}.

The differential cross section, the various structure functions and
polarization observables are all calculated starting from the 
following set of transition matrix elements
\begin{equation}
\label{eq:matr}
m^{fi} _ \lambda = 
<J_{R}M_{R} (E_x) ;\vec{p}_{N}, \frac {1}  {2} s_{N}^{z} \mid J_{\lambda }(q)\mid
J_{i}M_{i}> \; \; \; (\lambda = 0, \pm1) \;
\end{equation}
where $\mid \! J_{i}M_{i}\! >$ and $\mid \! J_{R}M_{R} (E_x)\! >$
refer to the quantum states of the target and residual A-1 nucleus and
$s_{N}^{z}$ denotes the spin projection of the ejectile along the
z-axis.  The latter is chosen to coincide with the direction of the
momentum transfer $\vec{q}$.  When calculating the induced and
transverse polarizations as defined in Eqs.~(\ref{eq:indu}) and
~(\ref{eq:trans}) the transition matrix element for the ejectile's
spin pointing in a certain direction determined by the polar and
azimuthal angle ($\theta ^*, \phi ^*$) are required.  These can be
easily obtained from the above matrix elements (\ref{eq:matr}) by
remarking that
\begin{eqnarray}
<J_{R}M_{R} & & (E_x) \vec{p}_{N}, \frac {1} {2} s'_{N}(\theta ^* \phi ^*)  
\mid  J_{\lambda }(q)\mid
J_{i}M_{i}> = \\ \nonumber & & 
\sum _ {s_{N}^{z}} \left( {\cal D}
_{s_{N}^{z} , s'_{N}} ^{(1/2)} \left( \theta ^*, \phi ^* \right) \right) ^* 
<J_{R}M_{R} (E_x) ;\vec{p}_{N}, \frac {1} {2} s_{N}^{z} \mid J_{\lambda }(q)\mid
J_{i}M_{i}> \; \; \; (\lambda = 0, \pm1) \; ,
\end{eqnarray}
where ${\cal D}^{(1/2)}$ is the Wigner ${\cal D}$-matrix for $j= \frac {1} {2}$ 
\begin{equation}
{\cal D}  ^{(1/2)} (\theta^*,\phi^*) =
\left( \begin{array} {cc} 
cos \frac {\theta ^*}{2} & sin \frac {\theta ^*}{2} e^{i \phi ^{*}} \\
sin \frac {\theta ^*}{2} & -cos \frac {\theta ^*}{2} e^{i \phi ^{*}} \\ 
\end{array} \right) \; .
\end{equation}

In determining the matrix elements of Eq.~(\ref{eq:matr}), a standard
multipole expansion for the electromagnetic current operators and the
``distorted'' outgoing nucleon wave is made.  The multipole expansion
of the current operators tends to converge more slowly as more
extended systems and higher momentum transfers are addressed.  For the
highest momentum transfer considered here ($\mid \vec{q} \mid \approx
1$~GeV), convergence in the $^{16}$O calculations could only be reached
after including all multipolarities up to $J_{max}$=30.  This number
is compatible with the estimate one would obtain by setting
$J_{max}=2qR_A$, where $R_A$ is the nuclear radius determined by $R_A
= 1.2 \; A^{1/3} \; \mathrm{(fm)}$.  The large number of multipoles
required for calculations at high momentum transfer is a serious
numerical complication that prevents models that were developed for
lower energy and momentum transfers from being easily extended to higher
energy and momentum domains.  Apart from the fact that a full
treatment of the relativistic effects would be in order, the
partial-wave expansion technique hampers extending the presented
formalism beyond the range of momentum-transfers considered here
($\mid \vec{q} \mid \le$1~GeV/c).

A quasi-elastic ($e,e'p$) process will predominantly excite those A-1 states $
\mid J_{R}M_{R} (E_x) >$ that bear a sizeable hole state component in
their overlap with the ground-state of the target nucleus. This
component of the final wave function can be constructed starting from the
standard multipole expansion in terms of particle-hole $\left|
p(lj\epsilon) h^{-1} \right>$ excitations out of the ground state of
the target nucleus \cite{Ryc88,Mah} :
\begin{eqnarray}
\left| J_{R}M_{R} (E_x) ;\vec{p}_{N}, \frac {1}  {2} s_{N}^{z}
\right> & = & \sum_{lm_ljm} \sum_{JM} 4 \pi i^l 
\sqrt{\frac {\pi} {2 \mu p_N}} <j_h \; m_h \; j\; m \mid J\; M> \nonumber \\
              & & \times <l \; m_l \; \frac{1}{2} \; s_{N}^{z} \mid
j \; m> e^{i(\delta_l+\sigma_l)}Y_{lm_l}^{*}(\Omega_N)
\mid p\left(lj\epsilon \right)h^{-1} \; ; \; JM > \;,
\end{eqnarray}
where $\epsilon \equiv p^2_N/(2 \mu)$, $\mu$ is the reduced mass of
the outgoing nucleon, $\delta _l$ is the central phase shift and
$\sigma _l$ is the Coulomb phase shift related to the electromagnetic
part of the mean-field potential.  The above expression has been
derived for the following conventions with respect to the asymptotic
behaviour of the continuum eigenstates $\mid p(lj\epsilon) > \equiv
\varphi_{lj}(r,E)$ of the mean-field potential~ and the total A-body
wave function:
\begin{eqnarray}
\varphi_{lj}(r,\epsilon) &  \stackrel{r \gg R_A} {\longrightarrow} &
\sqrt{\frac{2\mu} { \pi p_N}} \frac {sin(p_N r-\eta ln(2p_Nr) -
\frac{\pi l}{2} + \delta _l + \sigma _l)} {r} \; , \nonumber \\
\left| J_{R}M_{R} (E_x) ;\vec{p}_{N}, \frac {1}  {2} s_{N}^{z}
\right> &  \stackrel{r_A \gg R_A} {\longrightarrow} & \frac {1}
{\sqrt{A}} \left( e^{i \vec{p}_N \cdot \vec{r}_A} + f_k (\theta) 
\frac {e^{ip_Nr_{A}}} {r_{A}} \right) \left| \frac{1} {2} s_{N}^{z} \right>
 (-1)^{j_h+m_h} c_{h-m_{h}} \mid J_i M_i > \; , 
\end{eqnarray}
where $R_A$ is the nuclear radius.

\subsection{One and Two-body current operators}
\label{sec:current}

The one-body current is derived from the on-shell covariant
single-nucleon current
\begin{equation}
j^{\mu}= \overline u (\vec {p}_N,s_N^z) \Biggl[ F_1 (Q^2) \gamma ^{\mu} +
F_2 (Q^2) \frac {i \sigma ^{\mu \nu} q _{\nu}} {2 M_N} \Biggr]
u (\vec {p} _m,s_m^z) \; ,
\label{eq:dirac}
\end{equation}
where $F_1 (F_2)$ is the Dirac (Pauli) form factor.  Even when making
an abstraction from issues related to
off-shell corrections, the derivation of an appropriate current operator
with relativistic corrections to be used in calculations that adopt
a non-relativistic description for the strong interaction part of the
reaction process, is not
free from ambiguities \cite{amaro98,fearing94,boffi98,jeschonn98}.
For example, the two standard techniques to derive a non-relativistic
reduction of relativistic hamiltonians, i.e. the Foldy-Wouthuysen (FW)
and the ``direct Pauli reduction'' method, have been shown to make a
difference as far as the final expressions for the current operators
are concerned \cite{fearing94}.  Admittedly, the problems encountered
in implementing relativistic corrections in non-relativistic
calculations can be avoided by adopting a fully relativistic
description for the nuclear dynamics.  Over the years, a number of
fully relativistic models for ($e,e'p$) have been developed
\cite{wright,udias,johans,mcdermott}. Basically, all of these models 
start from a relativistic formulation of the mean-field idea and do
not embody current operators that go beyond the IA.
Unfortunately, for finite nuclei a realistic, empirically well-founded
and practicable relativistic nuclear-structure model that {\sl also accounts
for the multi-nucleon degrees of freedom} is as yet not available.
With this in mind we shall resort to a non-relativistic approach for
the strong interaction dynamics of the $(e,e'p)$ process.  We deem this
model to be a reliable and valuable testing ground to estimate the
relative importance of subnuclear degrees of freedom.  The
conventional way of deriving a nonrelativistic reduction of the above
one-body current operator is the FW technique which relies on a ($q/M_N$)
expansion.  Recently, an alternative expansion in terms of
$(p_m/M_N)$, with $p_m$ the momentum of the nucleon on which the
absorption occurs, has been suggested \cite{amaro98,jeschonn98}.
This expansion is based on the ``direct Pauli reduction'' method.  In
order $(q/M_N)^2$ and $(p_m/M_N)$ respectively, both methods give rise
to a nonrelativistic four-current density operator that reads in
coordinate space
\begin{eqnarray}  
\rho(\vec{r}) & = & \sum _{i=1} ^{A} \Biggl( C_1(q,Q^2) G_E^i(Q^2)  \delta (\vec{r}
- \vec{r} _i) 
\nonumber \\ & & - C_2(q,Q^2) \frac {(2 G_M^i(Q^2) - G_E^i(Q^2))} {8 M_N^2 i}
\vec{\nabla}  \cdot \left[ \vec{\sigma} _i \times    
\left\{ \vec{\nabla}_i, \delta (\vec{r} - \vec{r}_i) \right\}
\right] 
\Biggr) \nonumber \\
\vec{J}^{\perp}(\vec{r}) & = & \sum _{i=1} ^{A} C_3(q,Q^2) \Biggl(
\frac {G_E^i(Q^2)} {2M_N i} 
\left\{ \vec{\nabla}_i, \delta (\vec{r} - \vec{r}_i) \right\}
+ \frac {G_M^i(Q^2)} {2M_N} \delta (\vec{r} - \vec{r}_i) 
\left( \vec{\sigma}_i \times \vec{\nabla} \right)  
\Biggr) \; ,
\label{eq:redcurrent}
\end{eqnarray}
where $\{ \dots, \dots \}$ denotes the anticommutator and it is
implicitly understood that the operator $\vec{\nabla}$ acts only the
photon field.  The $G_E(Q^2)$ and $G_M(Q^2)$ are the Sachs form
factors for which we have adopted the standard dipole form.  The major
difference between the current operators in the two earlier sketched
approaches is not the operatorial form but the
derived expression for the coefficients ($C_1,C_2,C_3$).  Whereas with
the FW method one finds \cite{roccojoe}
\begin{equation}
\begin{array}{ccc}
C_1(q,Q^2) = \frac {1} {\sqrt{1+ \frac {Q^2} {4 M_N^2}}}
\hspace{0.7cm} & 
C_2(q,Q^2) = 1  \hspace{0.7cm} & C_3(q,Q^2) = 1 \hspace{0.7cm} \; ,\\
\end{array}  
\end{equation}
a $(p_m/M_N)$ expansion with the aid of the Pauli reduction technique
leads to \cite{amaro98,jeschonn98}
\begin{equation}
\begin{array}{ccc}
C_1(q,Q^2) = \frac {q} {\sqrt{Q^2}} \hspace{0.7cm} &  
C_2(q,Q^2) = \frac {1} {\sqrt{1+ \frac {Q^2} {4 M_N^2}}}
\hspace{0.7cm} & 
C_3(q,Q^2) = \frac {\sqrt{Q^2}} {q}  \hspace{0.7cm} \; .\\
\end{array} 
\end{equation}
Unless otherwise specified all results of this paper are obtained with
the four-current operator from Eq.(~\ref{eq:redcurrent}) using the $C$
coefficients as they are obtained with the FW method.  The major
difference in comparison with a strict non-relativistic approach is
the introduction of a spin-orbit term $\rho ^{so} (\vec{r})$ in the
charge density operator.  This gives rise to a spin-orbit term in the
Coulomb transition operator that determines the longitudinal strength
\begin{equation}
M_{JM}^{so}(q)= \int d \vec{r} j_J(qr) Y_{JM}(\Omega) \rho ^{so}
(\vec{r}) \; .
\end{equation}
In coordinate space the one-body transition matrix element
corresponding with this
operator reads
\begin{eqnarray}
& & \left< n_a l_a \frac {1} {2} j_a \parallel  M_{J}^{so}(q) 
\parallel  n_b l_b \frac {1} {2} j_b \right>  = 
\sum_{\eta \pm 1,J_2} (-1)^{\delta _{\eta,-1}+1}
C_2(q,Q^2) \frac {(2 G_M(Q^2) - G_E(Q^2))} {8 M_N^2} \nonumber \\   
& & \times 3 \sqrt{ 12(J+\delta_{\eta,+1})} (2J_2+1) 
\left\{ \begin{array}{ccc}
J_2 & J + \eta & 1 \\ 
1 & 1 & J \\
\end{array} \right\} 
\left\{ \begin{array}{ccc}
l_a & \frac {1} {2} & j_a \\ l_b & \frac {1} {2}  & j_b  \\
J_2 & 1 & J \\
\end{array} \right\} \nonumber \\
& & \times \left< n_a l_a \parallel 
\left( \frac {d} {dr} j_J(qr) + (-1)^{\delta_{\eta,+1}} (J + \delta
_{\eta,-1}) \frac {j_J(qr)} {r} \right) 
\left[ Y_{J+\eta} \otimes \left( \vec{\nabla} - \vec{\nabla}' \right)
\right]^{(J_{2})} \parallel n_b l_b \right> \; .
\end{eqnarray}

In determining the two-body current operators we start from the
observation that of all mesons carying the nucleon-nucleon interaction
the pions play a predominant role.  In our model calculations, the
current operators that explicitly account for subnuclear $\pi$ and
$\Delta$ degrees of freedom are effecively written in terms of the
coordinates of the two nucleons involved.  The meson-exchange and
$\Delta$-isobar currents as they were implemented in our calculations
have been discussed in detail in Ref.~\cite{jan97}.  The two-body
pion-exchange currents are derived from the one-pion exchange
potential in the standard fashion.  The derivation of the
$\Delta$-current operator is somewhat more complicated.  Indeed, this
operator cannot be constrained through charge-current conservation and
is therefore often referred to as a ``model-dependent'' operator
\cite{roccojoe}.  Moreover, of all pion-related two-body current
contributions the $\Delta$-current is the only one that could be
unambiguously shown to exhibit a quite strong medium dependence.
We do not consider explicit $\Delta$ admixtures in the wave functions, so that
all isobars in our model calculations are attached to a photon
line.  In our derivations, the $\pi N \Delta$ and $\gamma N \Delta$ coupling
are considered in the standard form~: 
\begin{eqnarray}
{\cal L} _{\pi N \Delta} & = & \frac {f_{\pi N \Delta}} {m _{\pi}}
\left( \vec{S} ^{\dagger} . \vec{\nabla} \right)
\left( \vec{T} ^{\dagger} . \vec{\pi} \right) \; ,
\\
{\cal L} _{\gamma N \Delta} & = & G_{\gamma N \Delta}(q_\mu^2)
\frac {f_{\gamma N \Delta}} {m _{\pi}}
\left( \vec{S} ^{\dagger} \times \vec{\nabla} \right) . \vec{A}
\; \vec{T} _{z} ^{\dagger} \; ,
\end{eqnarray}
where $\vec{S}$ and $\vec{T}$ denote the $\frac {1} {2} \rightarrow
\frac {3} {2}$ spin and isospin transition operators, $\vec{A}$ is the
external electromagnetic field and the coupling constants are $\frac
{f^2_{\pi N \Delta}} {4 \pi}$ = 0.37 and $f_{\gamma N\Delta}$ = 0.12.
The electromagnetic form factor of the delta $G_{\gamma N
\Delta}(q_\mu^2)$ is parametrized as \cite{marcucci98}
\begin{equation}
G_{\gamma N \Delta}(q_\mu^2)  = \frac {1} {\left( 1 - \frac {q_\mu^2}
{\Lambda_1 ^2} \right)^2}  
\frac {1} {\sqrt{ 1 - \frac {q_\mu^2}
{\Lambda_2 ^2} }} \; ,
\end{equation}
where $\Lambda_1$=0.84~GeV/c and $\Lambda_2$=1.2~GeV/c.  It is
interesting to note that the $N-\Delta$ electromagnetic form factor
$G_{\gamma N \Delta}$ is decreasing with $Q^2$ faster than the nucleon
dipole
form \cite{stoler93,frolov99}.  With the above coupling lagrangians we
arrive at the following expression for the $\Delta _{33}$-current in
momentum space
\begin{eqnarray}
\label{eq:pidelta}
&&\vec J_{\pi \Delta} (\vec q,\vec k_1,\vec k_2 ; \gamma N \Delta  \longrightarrow NN) =
{i \over 9}{{f_{\gamma N\Delta }f_{\pi NN}f_{\pi
N\Delta }} \over {m_\pi ^3\,}} G_{\gamma N \Delta}(q_\mu^2) \nonumber\\ 
&& \times \Biggl[ 
\left( G_{\Delta}^{I} + G_{\Delta}^{II} \right) 
\Biggl( 4  \vec \tau _{2,z} 
\left( \vec k_2 \times \vec q  \right) 
{{\vec \sigma _2\cdot \vec k_2} \over {\vec{k_2}^2+m_\pi
^2}} + 
4 {\vec \tau _{1,z}} 
\left( \vec k_1\times \vec q  \right) 
{{\vec \sigma _1\cdot \vec k_1} \over {\vec{k_1}^2+m_\pi ^2}}
\nonumber\\
&&  + (\vec \tau _1 \times \vec \tau _2)_z 
\left[
\left( {\vec \sigma _2 \times \vec k_1} \right) 
{{\vec \sigma _1\cdot \vec k_1} \over {\vec{k_1}^2+m_\pi ^2}} - 
\left( {\vec \sigma _1 \times \vec k_2} \right)
{{\vec \sigma _2 \cdot \vec k_2} \over {\vec{k_2}^2+m_\pi
^2}}  \right] \times \vec q  \Biggr) \nonumber
\\ && + \left( G_{\Delta}^{I} - G_{\Delta}^{II} \right)
\Biggl( - 2i  {\vec \tau _{2,z}}
 \left( \left( \vec \sigma _1 \times \vec k_2 \right) \times \vec q
\right) 
{{\vec \sigma _2 \cdot \vec k_2} \over {\vec{k_2}^2+m_\pi ^2}}
 -2 i \vec \tau _{1,z} 
\left( \left( \vec \sigma _2\times \vec k_1 \right) \times \vec q \right)
{{\vec \sigma _1\cdot \vec k_1} \over {\vec{k_1}^2+m_\pi ^2}} \nonumber\\
&& -2i (\vec \tau _1\times \vec \tau _2)_z 
\left[ \vec k_2 {{\vec \sigma _2\cdot \vec k_2} \over {\vec{
k_2}^2+m_\pi ^2}} -
\vec k_1 {{\vec \sigma _1\cdot \vec k_1} \over {\vec{
k_1}^2+m_\pi ^2}}  \right] \times \vec q  \Biggr)   \Biggr] \;.
\end{eqnarray} 
At the $\pi N \Delta$ and $\pi NN$ vertices, monopole form factors
\begin{equation}
\frac {\Lambda _ {\pi NN} ^2 - m _ \pi ^2} {\Lambda _ {\pi NN} ^2 + p _ \pi ^2}  
\end{equation}
are introduced.  They correct for finite size effects of the
interacting baryons and regularize the $\pi NN$ interaction at
extremely short distances.  All results presented below are obtained
with a cut-off parameter $\Lambda _{\pi NN}$ of 1250~MeV/c.  This value
corresponds with those that are typically obtained in the latest
parametrizations of the Bonn potential. In the above expression,
$G_{\Delta}^{I}$ ($G_{\Delta}^{II}$) denotes the propagator for a
$\Delta$ resonance that is created after (before) photoabsorption.  In
the calculations we adopt the following propagators
\begin{eqnarray}
G_{\Delta}^{I} & = & {{1} 
                    \over
                    {-\sqrt{s_{\Delta}^I} + M_{\Delta} - {i \over 2} \Gamma
                     _{\Delta}^{res}+ V_{\Delta}}} \nonumber \\
\label{eq:delta1}
G_{\Delta}^{II} & = & {{1} 
                    \over
                    {-\sqrt{s_{\Delta}^{II}} + M_{\Delta} }} \; ,
\label{eq:delta2}
\end{eqnarray}
where $\sqrt{s_{\Delta}}$ is the intrinsically available energy for the
resonance, $M_{\Delta}$=1232~MeV and $\Gamma _{\Delta}^{res}$ the
``free'' $\pi N$ decay width
\cite{Oset,dekker} 
\begin{equation}
\Gamma _{\Delta}^{res}  = \frac {2} {3} \frac {f _{\pi N \Delta} ^2} {4
\pi} \frac { \mid \vec{p} _{\pi} \mid ^3 } {m _{\pi} ^2}
\frac {M_N} {\sqrt{s_{\Delta}}} \; ,
\end{equation}
where $\vec{p} _{\pi}$ is the decay momentum in the center-of-mass
(c.o.m) frame of the $\pi N$ system.   

As we are dealing with off-shell nucleons the photon energy is not
completely available for internal excitation of the $\Delta_{33}$
resonance.  This effect is partially responsible for the observed
(real) energy shift of the $\Delta$ in the medium.  A reasonable
substitution for $s_{\Delta}$ is \cite{thomas,reply}
\begin{eqnarray}
\left( s_{\Delta}^I  \right) ^2 & = & -q_\mu q^{\mu} + 
      \left(M_N - \epsilon _h \right)^2 
+ 2 \omega (M_N - \epsilon _h) \\   
\left( s_{\Delta}^{II} \right)^2  & = & \left( \sqrt{
M_N^2  + \mid \vec q \mid ^2  } 
      - \omega  \right)^2 \; , 
\end{eqnarray}
where $\epsilon _h$ is the binding energy of the mean-field orbit on
which the pion is reabsorbed. The above expression accounts for the
observation that the delta peak in ($e,e'$) spectra 
section shifts to higher $\omega$'s with increasing momentum transfer.

\begin{figure}
\centerline{\epsfysize=10.cm \epsfbox{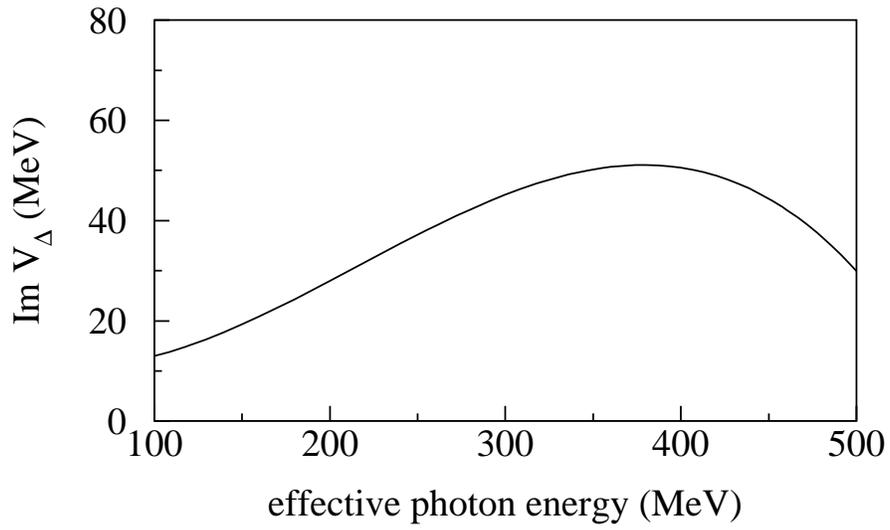}}
\caption{The imaginary part of the $\Delta$-medium potential as a
function of the effective photon energy (=$\omega - \frac{ q^{\mu}
q_{\mu}} {2M_N}$) for full nuclear density. The parametrization is
from Reference~\protect \cite{oset87}.}
\label{fig:width}
\end{figure}

Various pion-nucleus \cite{osterfeld}, real photoabsorption
\cite{bianchi} and electron scattering studies \cite{koch,oconnell}
experiments have pointed towards to strong medium modifications of the
$\Delta _{33}$ resonance.  For that reason, a medium-dependent term
$V_{\Delta}$ that accounts for the interaction of the $\Delta$
resonance with the nucleus was added to the propagator written in
(Eq.~\ref{eq:delta1}).  Various models that
implement a dynamical description of isobar propagation in the medium
have been developed \cite{koch,osterfeld}.  These models are generally
reasonably successful in describing the cross sections for inclusive
($e,e'$) reactions in the $\Delta _{33}$ resonance region
\cite{oconnell,cenni84}. The medium modifications of the
$\Delta$-resonance appear to be better under control in
electromagnetically induced processes than for example in
pion-absorption reactions \cite{korfgen94}.  For example, Chen and Lee
\cite{chen} have shown that a fairly good description of the
$^{12}$C($e,e'$) cross sections could be achieved with a $\Delta$
propagator of the type as written in Eq.~(\ref{eq:delta1}) provided
that one introduces a simple medium correction of the type
$V_{\Delta} [\mathrm{MeV}] = -30-40 \; \mathrm{i}$.  Similar values
for the $\Delta$-mass shift and broadening have been found in other
theoretical approaches \cite{kondra94} and were deduced from recent
total photoabsorption measurements \cite{bianchi}. For the
calculations presented here, we have used the $\Delta$-medium
potential $V_{\Delta}$ as it was calculated by Oset and Salcedo
\cite{oset87}.  In Figure \ref{fig:width} the imaginary part of
$V_{\Delta}$ is shown.  In the resonance region, this calculation does
indeed reproduce the earlier quoted value for the broadening of the
resonance.  The above two-body current operators have been used to
calculate $^{12}$C($\gamma,pp$) and $^{12}$C($\gamma,pn$) total cross
sections in the resonance region.  Not only could the position and
width of the resonance be reasonably reproduced, also the
($\gamma,pp$)/($\gamma,pn$) ratio for the various shell-model orbits
agreed fairly well with experiment \cite{douglas98}.  These results
lend confidence in the model assumptions with respect to the $\Delta$
currents and propagators.

\begin{figure}
\centerline{\epsfysize=10.cm \epsfbox{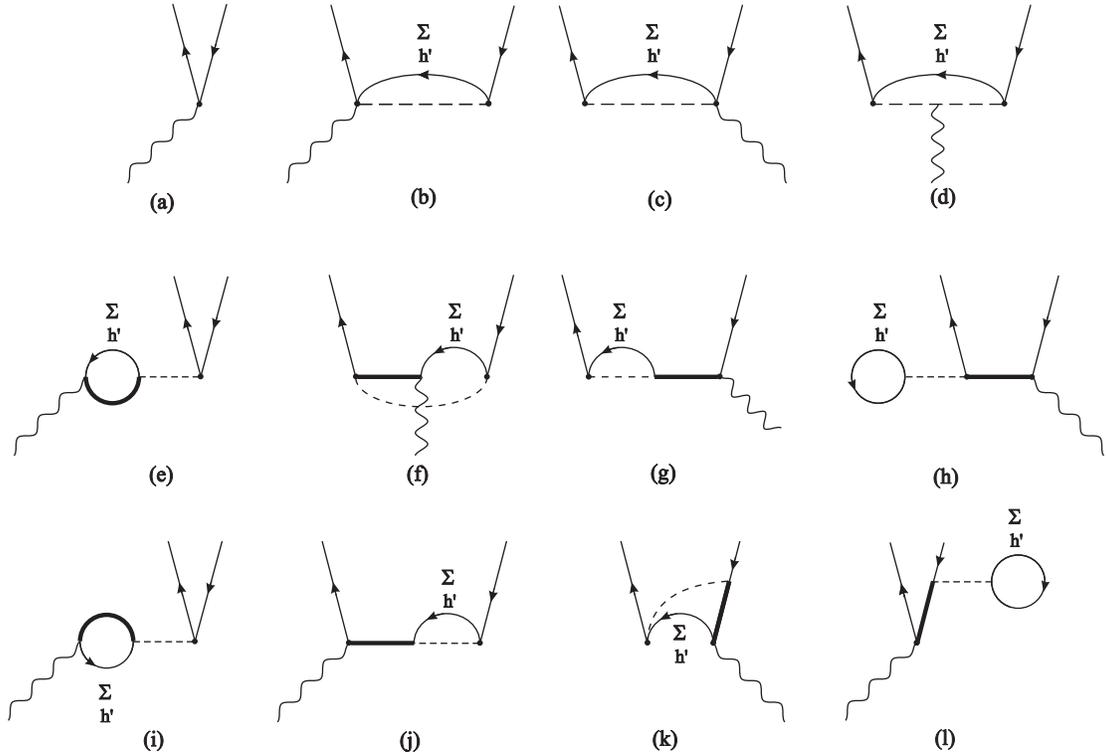}}
\caption{Diagrammatric representation of the reaction processes
included in the A($\vec{e},e'\vec{p}$) calculations.  Wavy, thin,
thick and dashed lines denote photons, nucleons, $\Delta$'s and pions.
Nucleon lines with an arrow down (up) refer to occupied (scattering)
states. Diagram (a) refers to the conventional IA. Diagrams (b)-(d) are related to
the MEC, whereas (e)-(l) represent the different direct and exchange IC contributions.}
\label{fig:diagram}
\end{figure}
\begin{figure}
\centerline{\epsfysize=14.cm \epsfbox{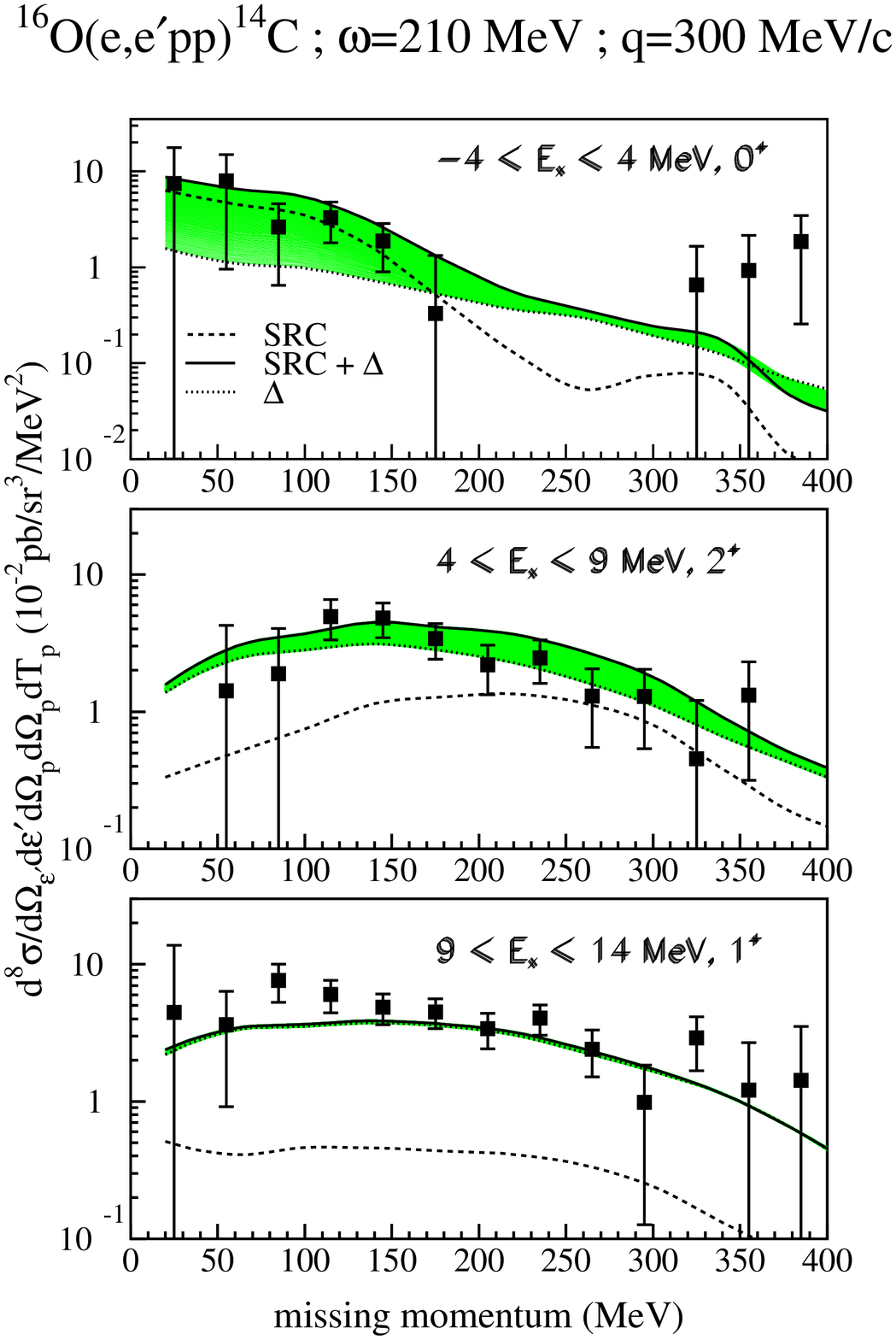} }
\caption{Calculated $^{16}$O($e,e'pp$) missing momentum distributions
for various groups of final states and electron kinematics determined
by $\epsilon$=580 MeV, $\epsilon'$=374 MeV and $\theta_e$=26.2$^o$.  The data and
curves refer to the two-proton knockout phase-space determined by the
polar angles 8$^o \le \theta_1 \le 40^o$, 115$^o \le \theta_2 \le
155^o$ and the kinetic energy of the backward going proton 52~MeV
$ \le $ T $_2  \le $ 108~MeV. The polar angles are expressed relative to the
direction of the momentum transfer. For the calculations, five mesh
points were considered in each of these variables and phase-space
averaging was pursued.  The data are from Ref.~\protect \cite{gercoprl2}. The
solid line is the result of the distorted-wave calculation when both
the SRC and $\Delta$ isobar effects are included.  The dotted (dashed)
line includes solely the $\Delta$ isobar (SRC) effects.}
\label{fig:gercoeep}
\end{figure}

Introducing the earlier discussed two-body current operators in
exclusive single-nucleon knockout calculations one ends with the
diagrams sketched in Figure~\ref{fig:diagram}.  The diagrams of
Figure \ref{fig:diagram}(b),(c) and (d) are the pion-exchange
contributions to single-nucleon knockout and necessarily imply a
charge-exchange mechanism carried by a charged pion.  As a
consequence, for proton knockout the sum over all occupied
single-particle states in the target nucleus ($\sum _{h'}$) produces
solely non-zero contributions for the neutron states.  The diagrams
from Figure~\ref{fig:diagram}(e)-(h) involve $\Delta _{33}$ excitation
after interaction of one of the target's nucleons with the (virtual)
photon field.  For this class of diagrams both charged and neutral
pion exchange belongs to the possibilities.  The contributions drawn
in Figures~\ref{fig:diagram}(e) and (h) involve exclusively neutral
pion exchange. Finally, the type of processes drawn in
Figure~\ref{fig:diagram}(i)-(l) involve a pre-formed $\Delta$ that is
deexcited after interacting with the photon field.  Technically
speaking, the introduction of the two-body current operators implies
that for each combination of a multipole component of the
electromagnetic transition operator $T_{J}$ and a partial wave of the
ejectile's wave function ($\mid p(lj\epsilon) >$), two-body matrix
elements of the  type
\begin{eqnarray}
\label{eq:two}
& & <p(lj\epsilon) h^{-1} ; \; J  \mid\mid  T_{J}^{[2]}(q)\mid\mid 0^{+}(g.s.)>
 =  \sum_{h' J_{1}J_{2}} \sqrt{2J_{1}+1}\sqrt{2J_{2}+1}(-1)^{j_{h}-j_{h'}
-J-J_{2}}
\left\{ \begin{array}{ccc}j_{h} &j_{h'}& J_{1}\\J_{2} &J& j \end{array}
\right\}\nonumber\\
&&\times \Bigl( <h h' \; ; \; J_{1} \mid\mid T_{J}^{[2]}(q)\mid\mid p(lj
\epsilon) h' \; ; \; J_{2}> - (-1)^{j'_h+j+J_2}
<h h' \; ; \; J_{1} \mid\mid T_{J}^{[2]}(q)\mid\mid h' p(lj
\epsilon)  \; ; \; J_{2}> \Bigr) \; ,
\label{eq:mattwo}
\end{eqnarray}
are to be coherently added to the conventional
one-body current contribution from the Impulse Approximation (IA)
\begin{equation}
<p(lj\epsilon)  h^{-1} ; \; J  \parallel  T_{J}^{[1]}(q) \parallel  0^{+}(g.s.)>
\; .
\end{equation}
In the above expression, $T_{J}^{[1]}$ and $T_{J}^{[2]}$  is the one
and two-body current contribution to the
electromagnetic transition operator.  The explicit expressions for the
reduced two-body matrix elements with the meson-exchange and isobar
currents can be found in Refs.~\cite{jan97,jana568}.  The sum over
$h'$ involves all occupied proton and neutron single-particle states.
It speaks for itself that for a nucleus like $^{208}$Pb this summation
can only be performed at a large computational cost.  For consistency
reasons and to avoid orthogonality deficiencies, the wave functions
for all occupied states $h'$ are calculated in exactly the same
mean-field potential in which also the distorted outgoing nucleon wave
and the overlap wave function $\left<J_R M_R (E_x) \mid J_i M_i
\right>$ is determined.  We remark that the very same type of matrix
elements that dictate the two-body current contributions to the
single-nucleon knockout processes, determine the two-body current
contributions to the cross sections for the A($\gamma,NN$) and
A($e,e'NN$) processes that are presently the subject of investigation
at various laboratories.  It speaks for itself that two-nucleon
knockout processes represent an intrinsically superior way of
exploring the two-nucleon effects in nuclei. The model assumptions
with respect to the current operators adopted here, are indentical to
those that we have adopted in our two-nucleon knockout studies.  Also
the potential in which the bound and scattering states are calculated
are indentical for the single- and two-nucleon knockout studies.  To
illustrate the potential of two-nucleon knockout studies to acquire a
precise understanding of two-nucleon mechanisms in finite systems,
Figure~\ref{fig:gercoeep} shows a comparison of recently obtained
$^{16}$O($e,e'pp$) data and our model calculations for the three
lowest bins in the excitation energy spectrum in $^{14}$C.  The
comparison between the calculations and the data is done as a function
of the pair missing momentum $\mid \vec{P} \mid = \mid \vec{p}_1 +
\vec{p}_2 - \vec{q} \mid$, 
which is the c.m. momentum of the pair before it undergoes the
electromagnetic interaction with the photon field and is established
to be the scaling variable in two-nucleon knockout processes, i.e. the
counterpart of the variable $\mid \vec{p}_m \mid = \mid \vec{p}_N -
\vec{q} \mid$ in the ($e,e'p$). For the data shown in
Figure~\ref{fig:gercoeep} the experimental resolution in the
excitation (or missing) energy of $^{14}$C was of the order of 4 MeV
and the individual states could not be resolved in the missing energy
spectrum.  On the other hand, high-resolution $^{16}$O($e,e'pp$)
investigations from Mainz \cite{guenther} and
$^{15}$N($d,^3He$)$^{14}$C transfer reactions \cite{kaschl71} allow to
infer that the lowest excitation-energy bin (-4~MeV $\leq E_x \leq
$4~MeV) is exclusively fed through the $^{14}$C ground-state
transition, whereas the second (4~MeV $\leq E_x \leq $9~MeV) and third
bin (9~MeV $\leq E_x \leq $14~MeV) are mainly fed through the $\left|
2^+ ; E_{x} = 7.01,8.32~\mathrm{MeV} \right>$ doublet and the $\left|
1^+ ; E_{x}=11.3~\mathrm{MeV} \right>$ state respectively.  These
states can be safely quoted to be the only ones in the low
excitation-energy spectrum of $^{14}$C that have a dominant
``two-hole'' structure relative to the ground state of $^{16}$O.  This
information allows to constrain the nuclear-structure input that is
required for the $(e,e'pp)$ calculations. The overall agreement
between the calculations, that are parameter-free, and the data is
reasonable and lends support for the model assumptions with respect to
the $\Delta$-current operator and propagator.  Indeed, for the
transitions to the 2$^+$ and $1^+$ states intermediate $\Delta$
creation is predicted to be the dominant reaction process in the
($e,e'pp$) reaction under evaluation. The shaded region in
Figure~\ref{fig:gercoeep} is the calculated contribution from the
central Jastrow correlations to the two-proton knockout cross
sections.  They make a big contribution to the ground-state transition
(upper panel) in the low missing momentum regime $P \leq 250$~MeV/c
which can be inferred to arise from diproton knockout from heavily
correlated $^{1}S_0$(T=0) pairs \cite{gercoprl2}.  For the results presented in
Figure~\ref{fig:gercoeep} the contribution from the Jastrow or
short-range correlations (SRC) was calculated with a correlation
function as it was obtained from a G-matrix calculation by Gearhart
and Dickhoff \cite{gearhart}. Recently, this correlation function was
shown to produce a favorable agreement with $^{12}$C($e,e'pp$) data
\cite{blom98}. In comparison with other model predictions for the
central correlation function, the one obtained by Gearhart and Dickhoff
should be classified in between the categories of ``hard'' (with a
core at short internucleon distances) and ``soft'' (characterized by a
finite probability to observe nucleon pairs at very short internucleon
distances).

\section{Results and discussion}
\label{sec:results}

\subsection{$^{16}$O($\vec{e},e'\vec{p}$) and
$^{12}$C($\vec{e},e'\vec{p}$) results for 0.05~$\leq Q^2$~$\leq$0.50~(GeV/c)$^2$} 
\begin{figure}
\begin{center}
\setlength{\unitlength}{1cm}
\begin{picture}(16,20)
\put(1.00,8.0){\mbox{\epsfysize=9.5cm\epsffile{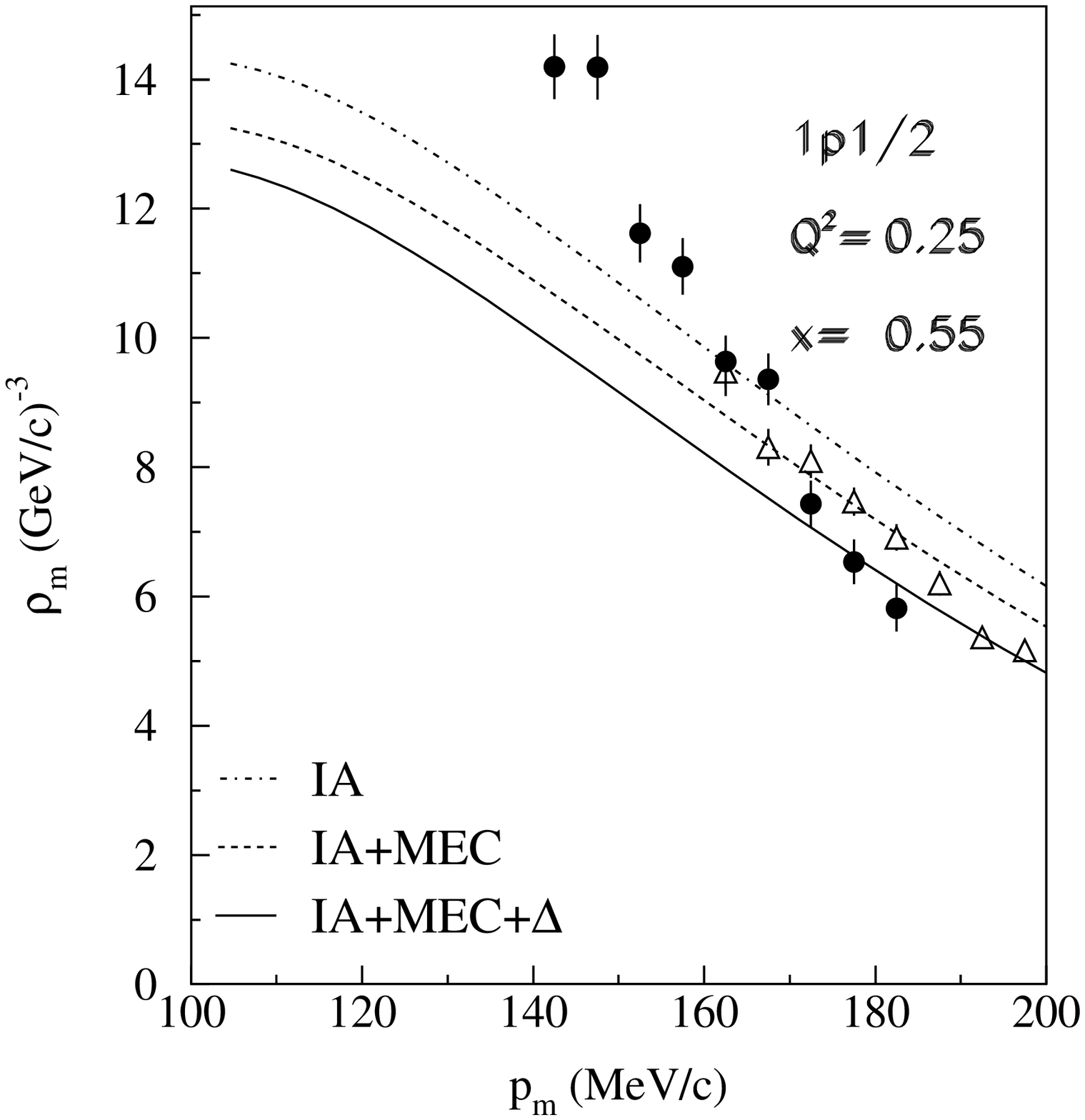}}}
\put(8.5,8.0){\mbox{\epsfysize=9.5cm\epsffile{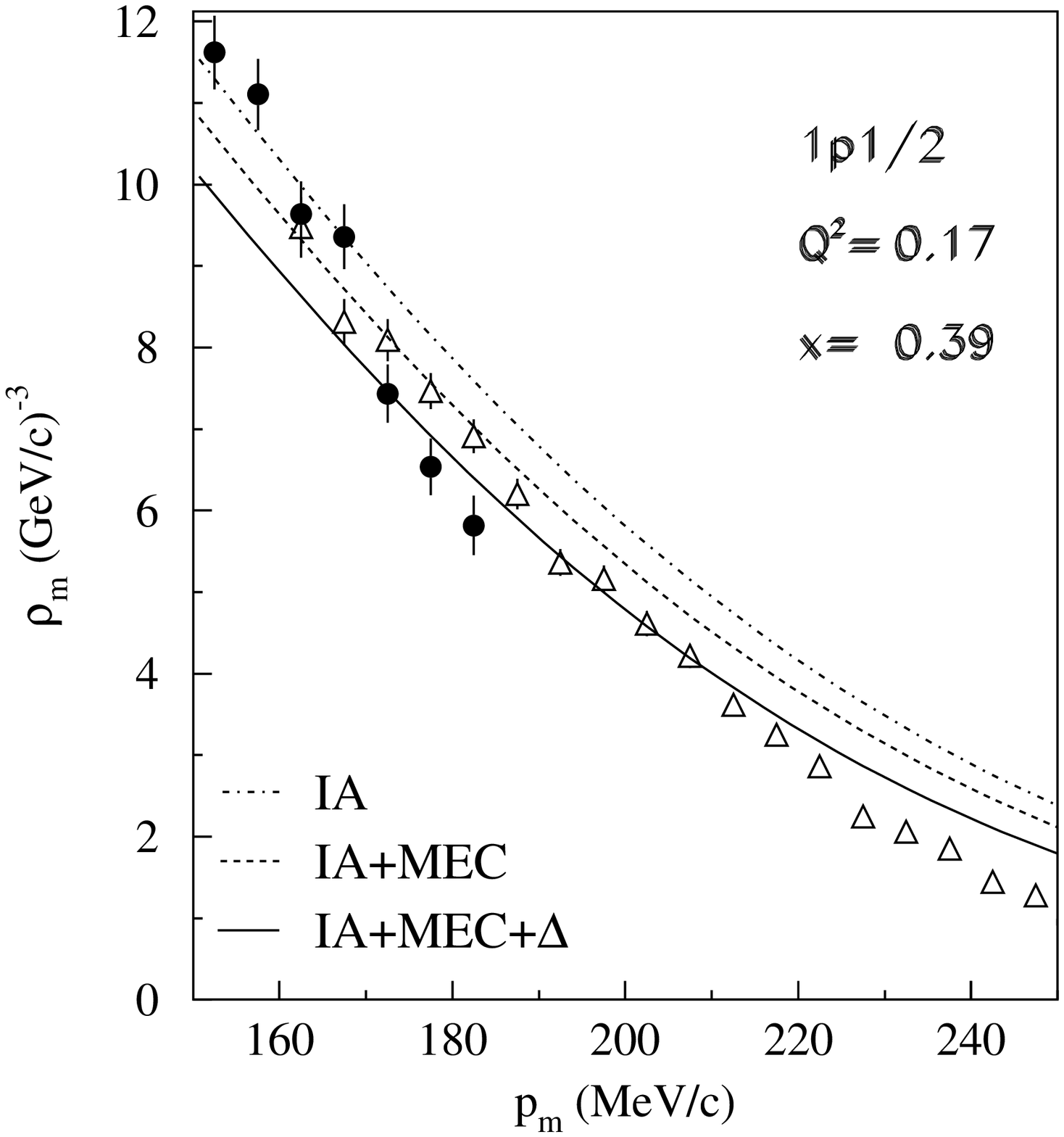}}}
\put(1.00,0.0){\mbox{\epsfysize=9.5cm\epsffile{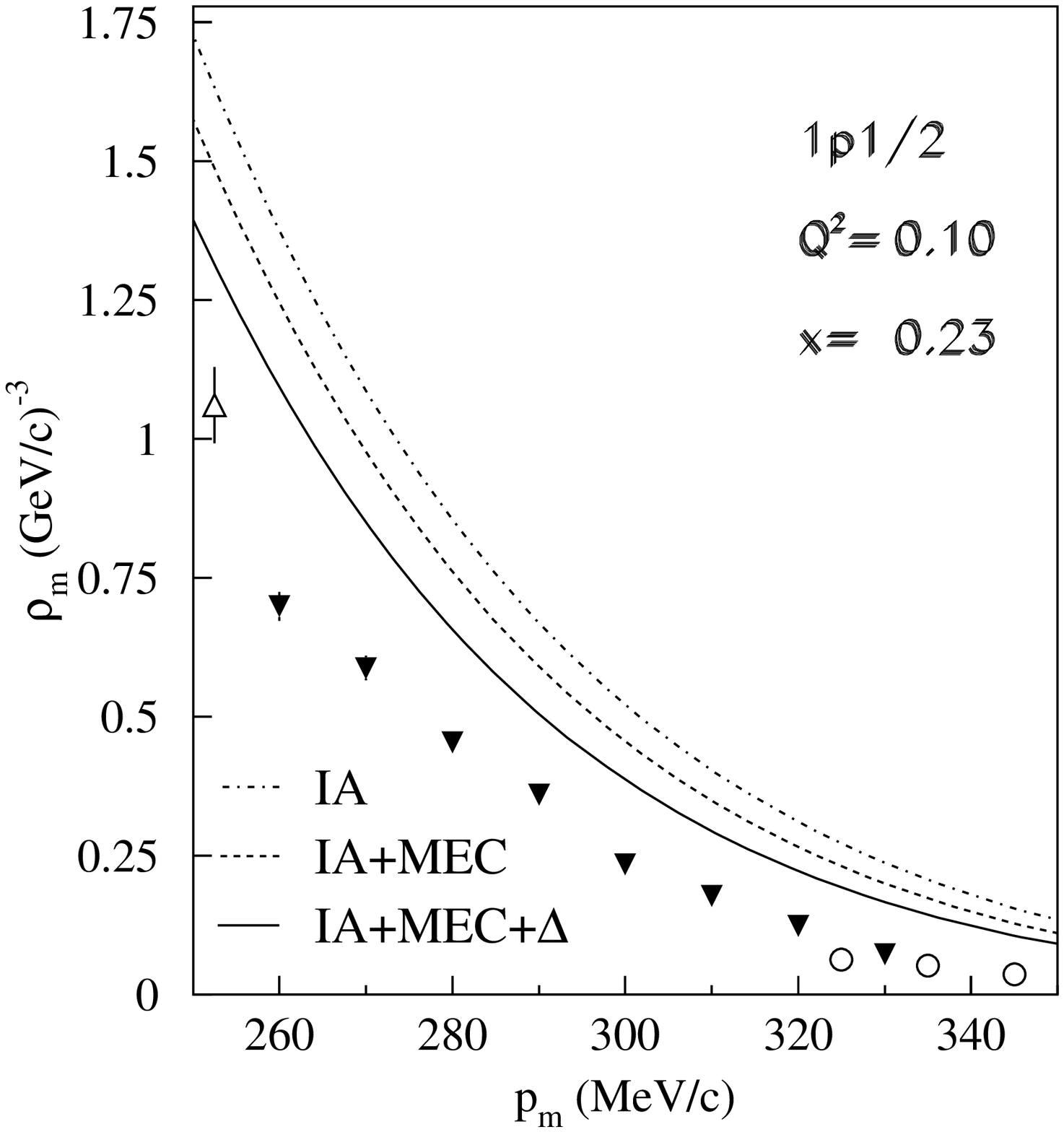}}}
\put(8.5,0.0){\mbox{\epsfysize=9.5cm\epsffile{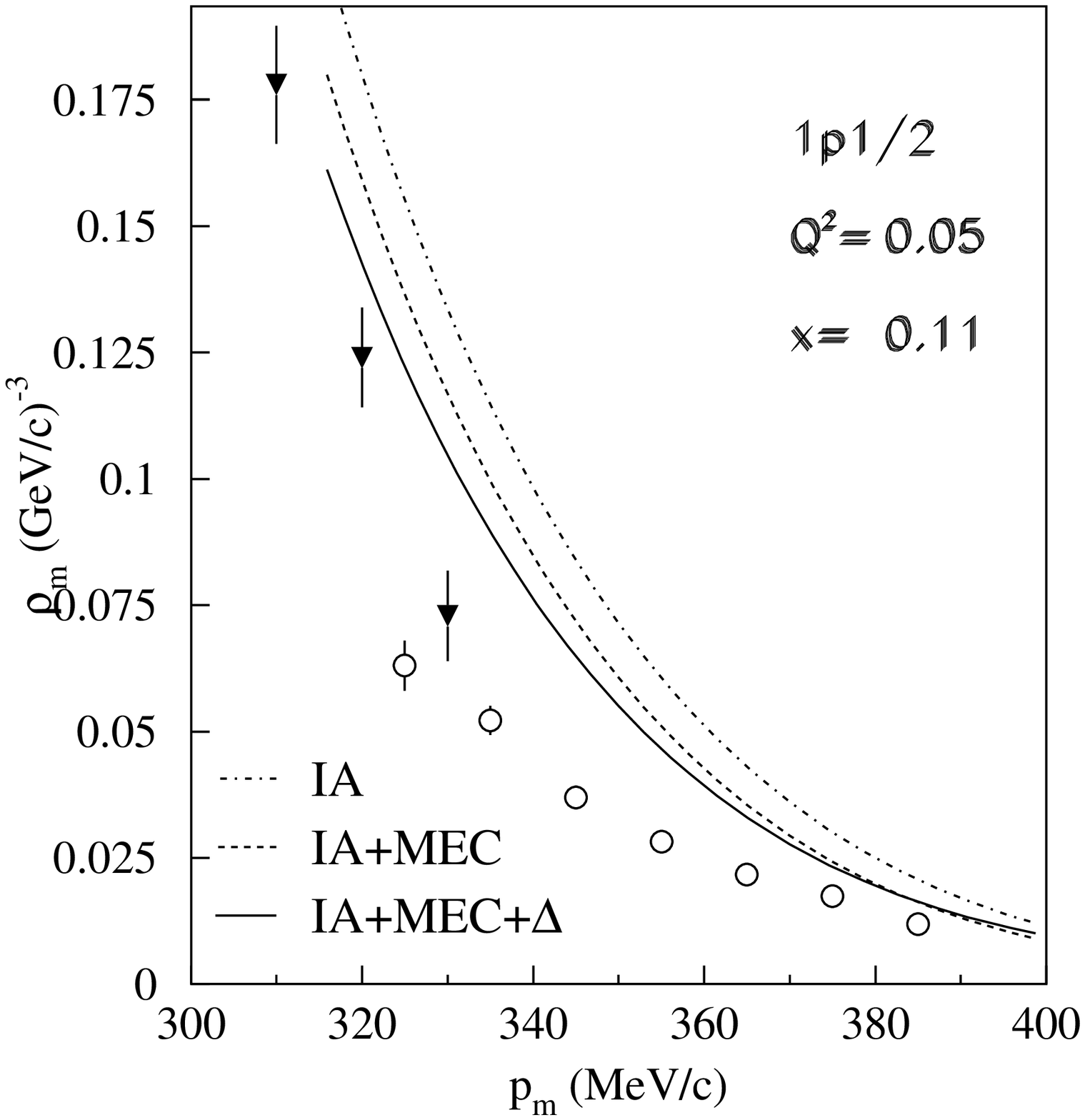}}}
\end{picture}
\end{center}
\caption{Reduced cross sections for the
$^{16}$O($e,e'p$)$^{15}$N(1/2$^-$,g.s.) reaction in parallel
kinematics.  The initial electron energy is kept constant at
$\epsilon$=855.1~MeV.  Starting from the upper left panel and going
clockwise the momentum transfer q is respectively 0.55,0.47,0.39 and
0.3~GeV/c.  The dot-dashed curve shows the result for the impulse
approximation ; in the dashed curve MEC effects are also included, and
the solid curve represents the full calculation including also IC. For
each case a central value of $Q^2$ and Bjorken $x$ is given.}
\label{fig:blom1}
\end{figure}

\begin{figure}
\begin{center}
\setlength{\unitlength}{1cm}
\begin{picture}(16,20)
\put(1.00,8.0){\mbox{\epsfysize=9.5cm\epsffile{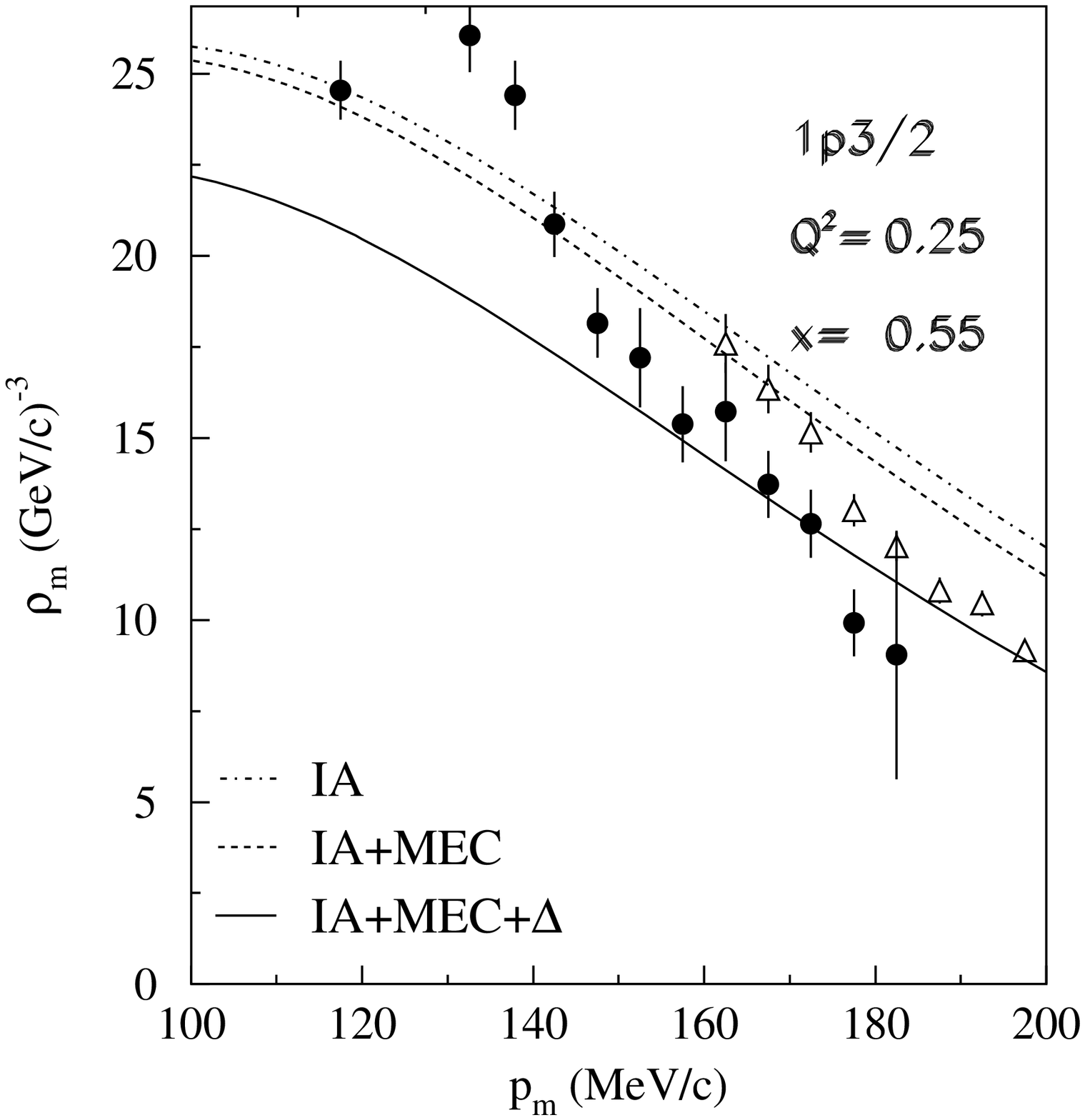}}}
\put(8.5,8.0){\mbox{\epsfysize=9.5cm\epsffile{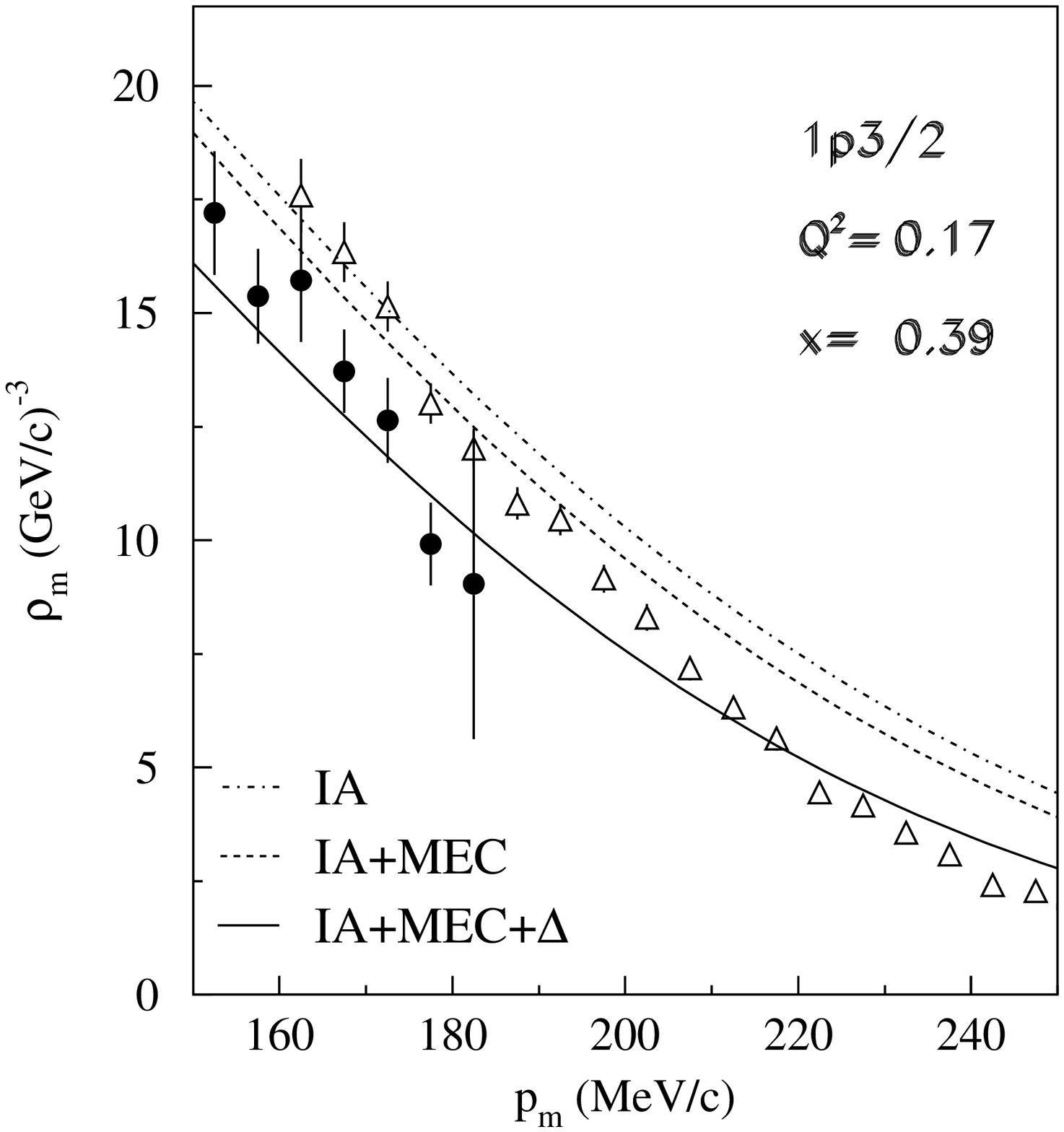}}}
\put(1.00,0.0){\mbox{\epsfysize=9.5cm\epsffile{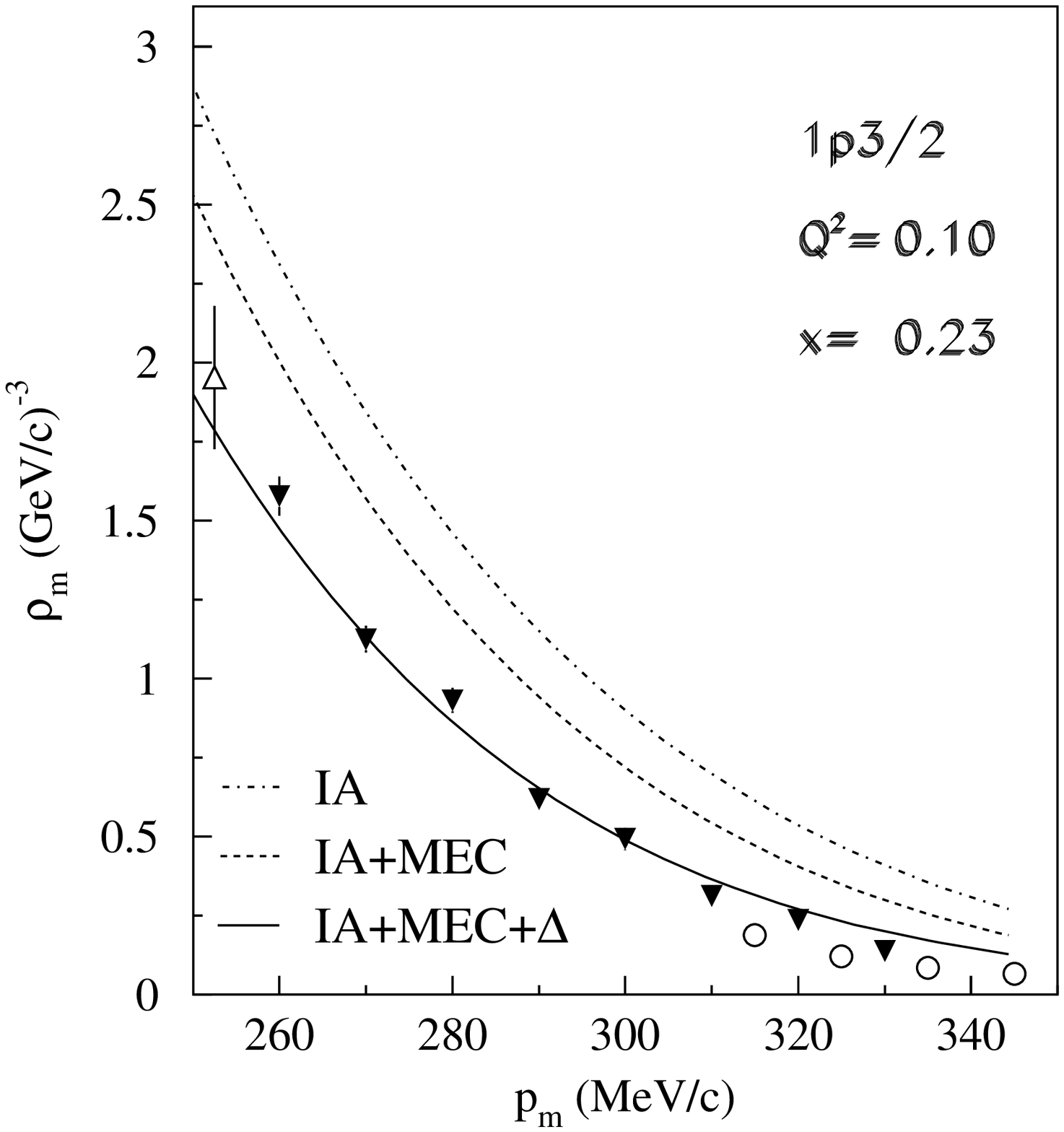}}}
\put(8.5,0.0){\mbox{\epsfysize=9.5cm\epsffile{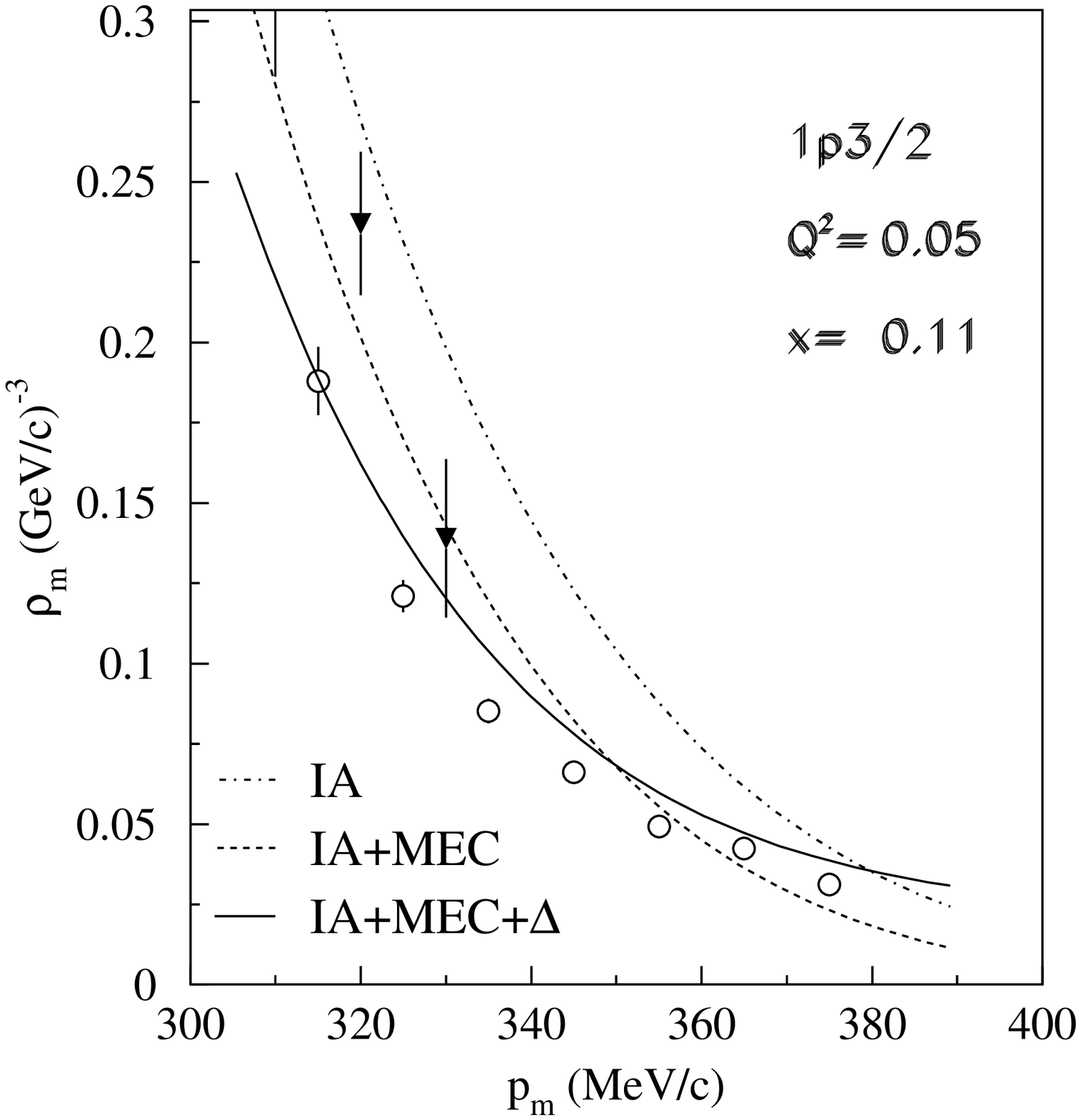}}}
\end{picture}
\end{center}
\caption{As in Figure~\ref{fig:blom1}, but now for the
$^{16}$O($e,e'p$)$^{15}$N(3/2$^-$,6.32~MeV) transition.}
\label{fig:blom2}
\end{figure}

We start our investigations into the role of two-currents for the
kinematics of an $^{16}$O($e,e'p$) experiment made at MAINZ and
reported in Reference~\cite{blomqvist95}.  In this experiment, the
cross sections were measured in a broad missing momentum range between
100 and 700~MeV/c.  The data were compared to distorted wave impulse
approximation (DWIA) calculations.  At
lower missing momenta ($p_m~\leq$300~MeV/c) the data were observed to
overshoot these calculations by a factor of two, whereas for the highest
missing momenta probed the opposite effect was noticed.  This
qualitative behaviour was recently confirmed in an independent
calculation by J.J. Kelly \cite{kelly1}. Despite the fact that rather
large bound nucleon momenta were probed it was concluded that the
deviations from standard (Woods-Saxon) mean-field wave functions are
modest.  This observation lends support for the picture that
nucleon-nucleon correlations are related to the hard-core part of the
nucleon-nucleon interaction and are not expected to bring about large
deviations from mean-field models in single-knockout processes as long
as low excitation energies in the A-1 system are probed.  The
data of Ref.~\cite{blomqvist95} were obtained for an initial electron
energy of 855.1~MeV and $T_p \approx$195~MeV. As a large fraction of
the data were taken in parallel kinematics ($\vec{q} \; \| \; \vec{p}_N$), the
momentum transfer had to be decreased as higher missing momenta were
probed.  Consequently, to reach higher missing momenta one had to move
out of quasi-elastic kinematics. Figures~\ref{fig:blom1} and
\ref{fig:blom2} show the predicted sensitivity of the Mainz results to
the two-body currents.  The Figures show results for four different
central values in $Q^2$ and parallel kinematics.  The kinematics is
such that one moves out of quasi-elastic conditions as higher missing
momenta are probed.  As is commonly done, we
present the angular cross-sections in terms of the reduced
cross section ``$\rho _m$''.  Hereby, we adopt the standard convention
of dividing the calculated cross sections by a kinematical factor
times the so-called ``CC1'' \cite{forest} elementary electron-proton
cross section
\begin{equation}
\rho_m \equiv \frac 
{\frac {d^5 \sigma }  {d \Omega _N d \epsilon ' d \Omega
_{\epsilon '}}}  
{{\frac {p_N  E_N} {(2\pi)^3}}   f_{rec}^{-1} \sigma_{ep}^{CC1}} \; .
\end{equation}
 
Referring to Figures \ref{fig:blom1} and \ref{fig:blom2}, striking
features are: (1) the sensitivity of the cross sections to the
two-body currents is substantial and (2) the two-body currents do not
dramatically alter the shape of the reduced cross section for none of
the four $Q^2$ values considered.  As a consequence, the impact of the
two-body currents would generally not be noticed when comparing
results of calculations performed within the IA with data but would
simply be ``effectively'' accounted for in the spectroscopic factor
that is used to scale the DWIA calculations to the data.  The net
effect of the two-body currents is a reduction of the cross sections.
This reduction is of the order 10-20\% for the lowest missing momenta
range and can amount to a factor of two at $p_m \approx 300$~MeV/c.
Note that for the highest missing momenta considered in Figures
\ref{fig:blom1} and \ref{fig:blom2} inclusion of the two-body currents
tends to increase the cross section.  This is generally the case when
higher missing momenta are probed \cite{kester96}.  The observed
reduction at low missing momenta that was ascribed to two-body current
effects is not sufficient to explain why the data are substantially
lower (factor of two) than expected on the basis of DWIA models.
Indeed, the curves of Figures \ref{fig:blom1} and \ref{fig:blom2} are
obtained with a reduction factor of 0.25 for the ground-state
transition and 0.30 for the
$^{16}$O($e,e'p$)$^{15}$N(3/2$^-$,6.32~MeV) transition.  With these
values we obtain reasonable visual fits of the data.  The transparency
at the considered four-momentum transfers $Q^2$ is experimentally
determined to be about 0.75 for a light nucleus like $^{12}$C
\cite{potterfeld,garino92}.  When correcting the reduction factors we
have applied to obtain visual fits of the data, for transparency (or
attenuation) corrections that fall beyond the scope of our model
calculations, we obtain the spectroscopic factors $S_{lj}
(1/2^-,$E$_x$=0.0~MeV)$\approx$ 0.67 and
$S_{lj}(3/2^-,$E$_x$=6.3~MeV)$\approx$1.6.  In line with the
theoretical conclusions drawn from the analyses of
References~\cite{blomqvist95} and \cite{kelly2} these values are
substantially smaller than the values obtained with high-resolution
measurements at lower proton kinetic energies where it was found that
$S_{lj} (1/2^-,$E$_x$=0.0~MeV)=1.20 and
$S_{lj}(3/2^-,$E$_x$=6.3~MeV)=2.00 \cite{leuschne94}.  Note, however,
that for $1p_{3/2}$ knockout the spectroscopic factor that is deduced
from the calculations that include the two-body currents is about 30\%
larger than the value that would be deduced in the IA.

Another striking feature of the results displayed in Figures
\ref{fig:blom1} and \ref{fig:blom2} is that the influence of the
subnuclear degrees-of-freedom is more pronounced for knockout from the
``stretched'' ($j_h= l_h + \frac{1}{2}$) orbit (i.e. $1p_{3/2}$) than
from its ``jack-knifed'' spin-orbit partner ($j_h= l_h - \frac{1}{2}$)
(i.e. $1p_{1/2}$).  The major reason for this behaviour is simply that
all the two-body current operators have a strong spin dependency.  It
is worth mentioning that the relativistic effects arising from the
lower (or negative energy) components in the bound state wave
functions were recently shown to have a similar sort of sensitivity.  Indeed,
the (relativistic) $(e,e'p)$ results for knockout from the
``stretched'' orbits will be closer 
to their respective non-relativistic limits
than those for ``jack-knifed''
orbits  \cite{caba98}.
Accordingly, the predicted relativistic effects attributed to the small
components in the bound state
wave functions are smallest for these transitions with maximal
two-body current contributions.

\begin{figure}
\begin{center}
\setlength{\unitlength}{1cm}
\begin{picture}(16,12)
\put(-1.00,0.0){\mbox{\epsfysize=11.cm\epsffile{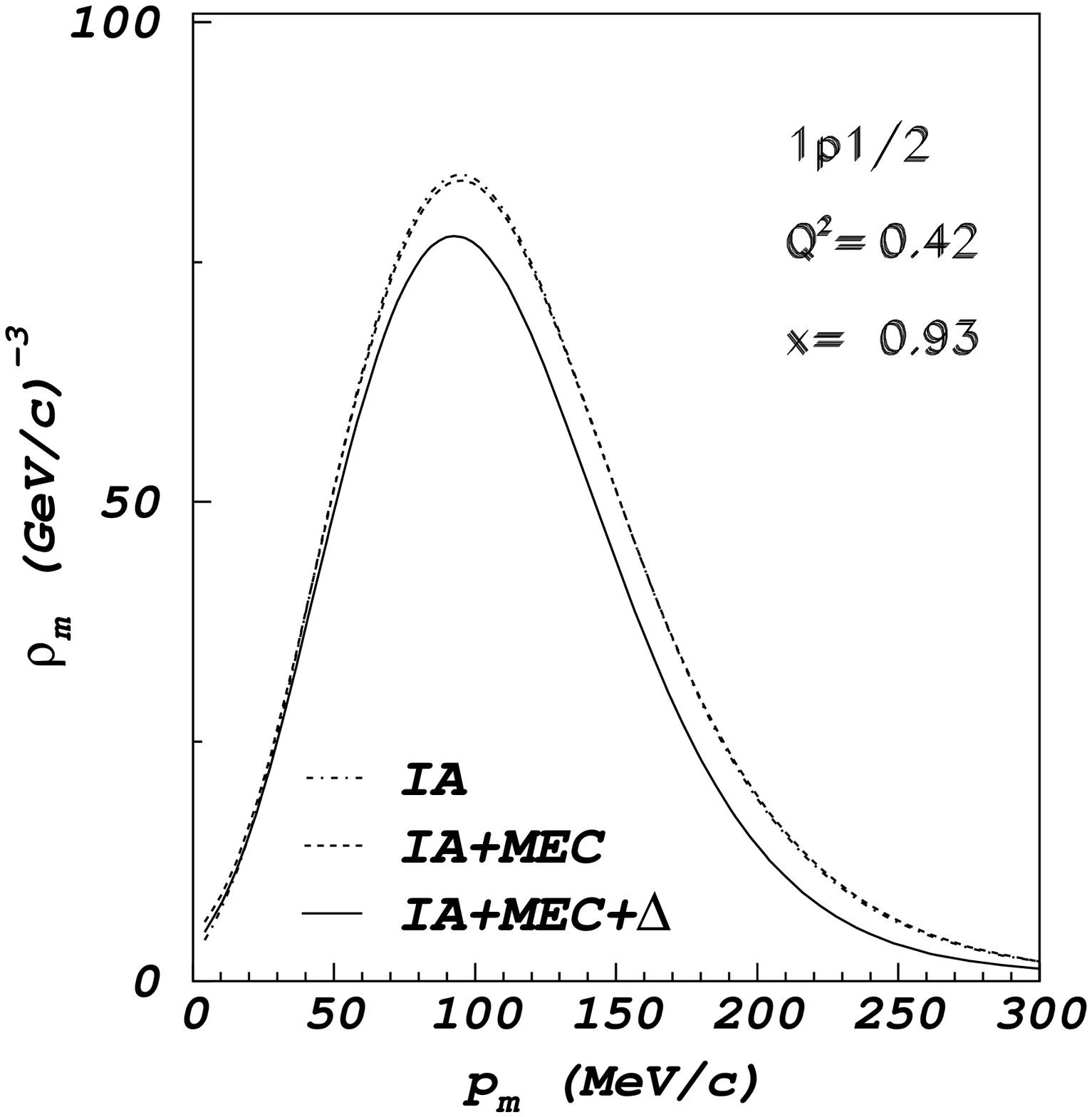}}}
\put(7.50,0.0){\mbox{\epsfysize=11.cm\epsffile{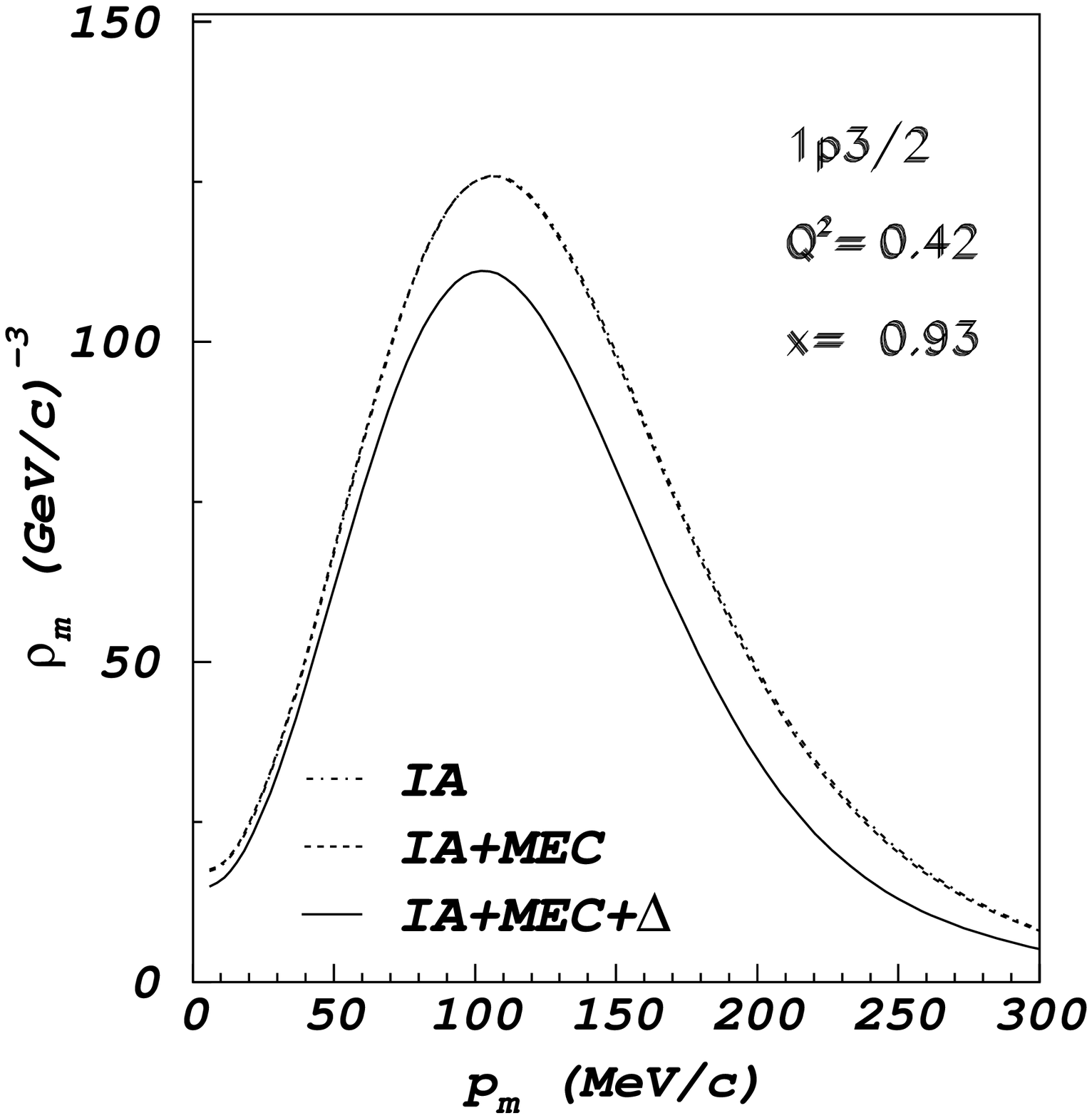}}}
\end{picture}
\end{center}
\caption{Reduced cross sections for $^{16}$O($e,e'p$) from
p-shell states in quasi-perpendicular kinematics.  The electron
kinematics is determined by $\epsilon$=0.855~GeV, $\omega$=240~MeV and
q=0.69~GeV/c.  The dot-dashed curve shows the result for the impulse
approximation ; in the dashed curve MEC effects are also included, and
the solid curve represents the full calculation
including also IC. The curves are normalized for full p-shell occupancy.}
\label{fig:mommainz}
\end{figure}

\begin{figure}
\begin{center}
{\mbox{\epsfysize=15.cm\epsffile{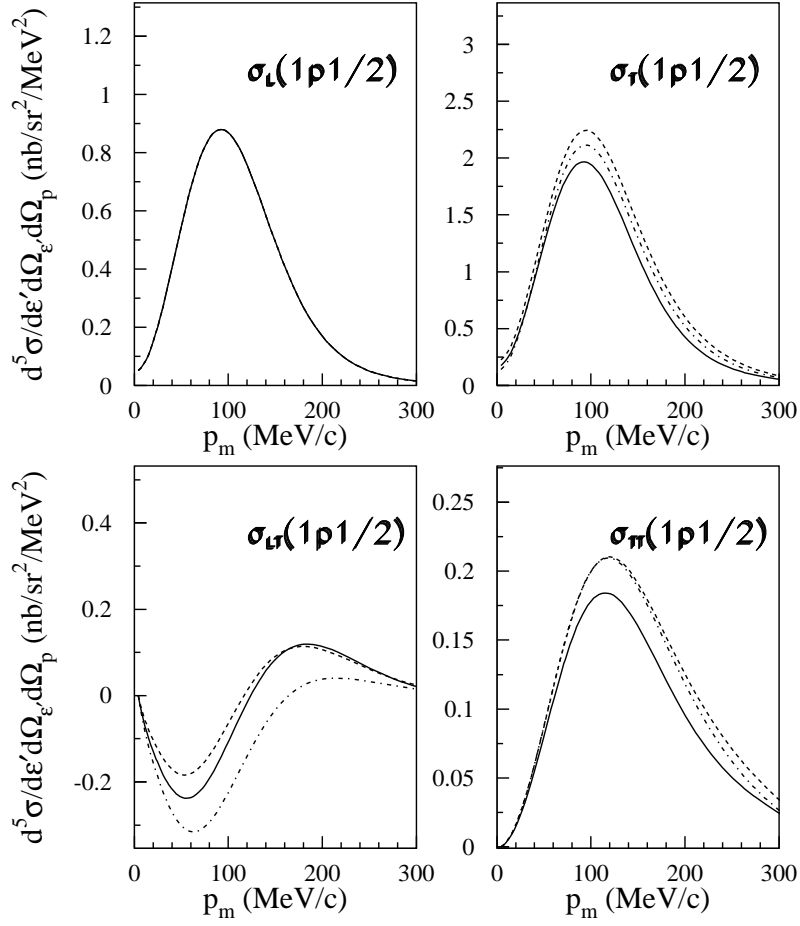}}}
\end{center}
\caption{The different contributions to the 
$^{16}$O($e,e'p$)$^{15}$N(1/2$^-$) cross section for the
kinematics of Figure~\ref{fig:mommainz}. The dot-dashed curve shows
the result for the impulse approximation ; in the dashed curve MEC
effects are also included, and the solid curve represents the full
calculation including also IC. The curves are normalized for full
$1p_{1/2}$ occupancy.}
\label{fig:strmainz}
\end{figure}

\begin{figure}
\begin{center}
\setlength{\unitlength}{1cm}
\begin{picture}(16,12)
\put(-1.00,0.0){\mbox{\epsfysize=12.cm\epsffile{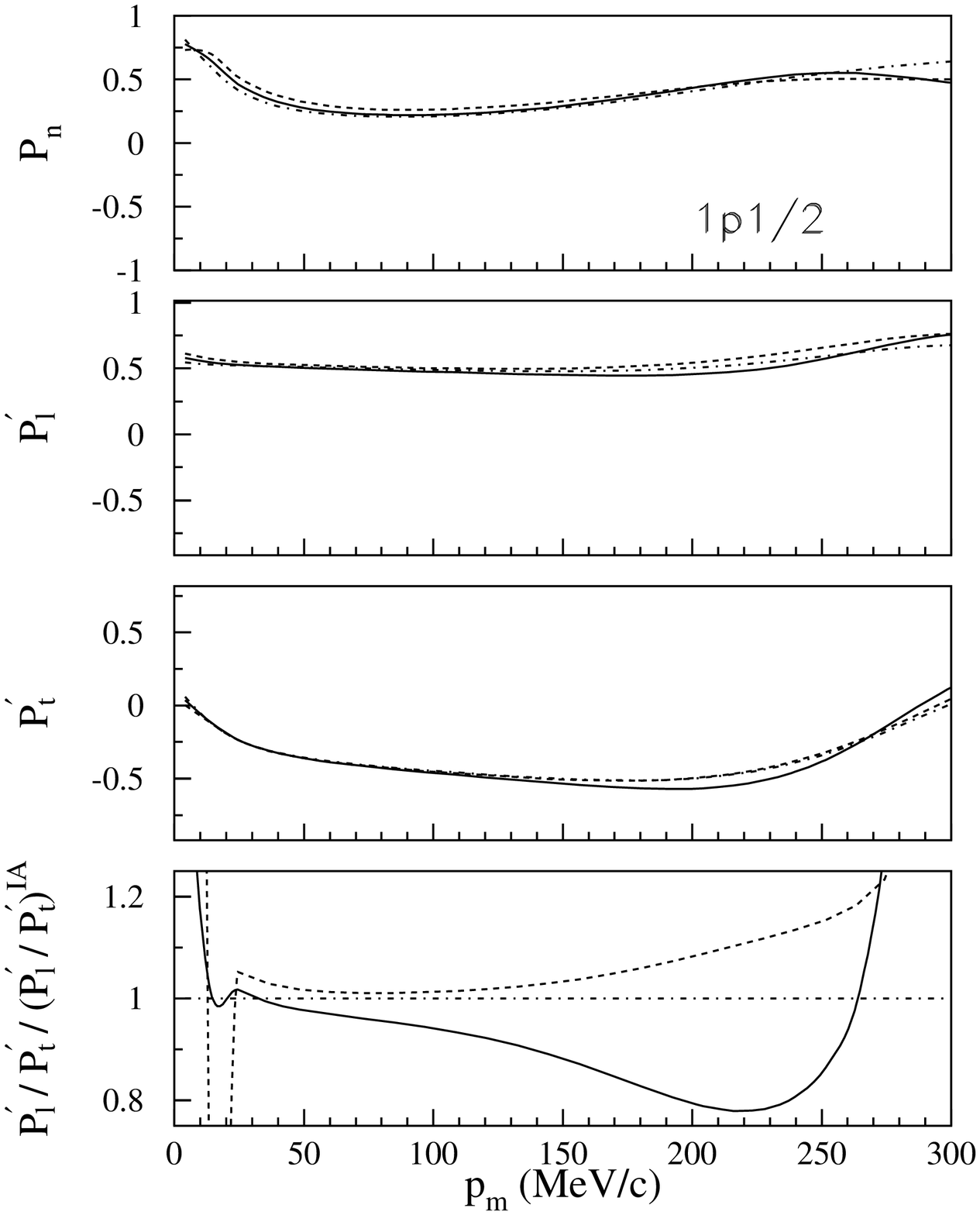}}}
\put(7.50,0.0){\mbox{\epsfysize=12.cm\epsffile{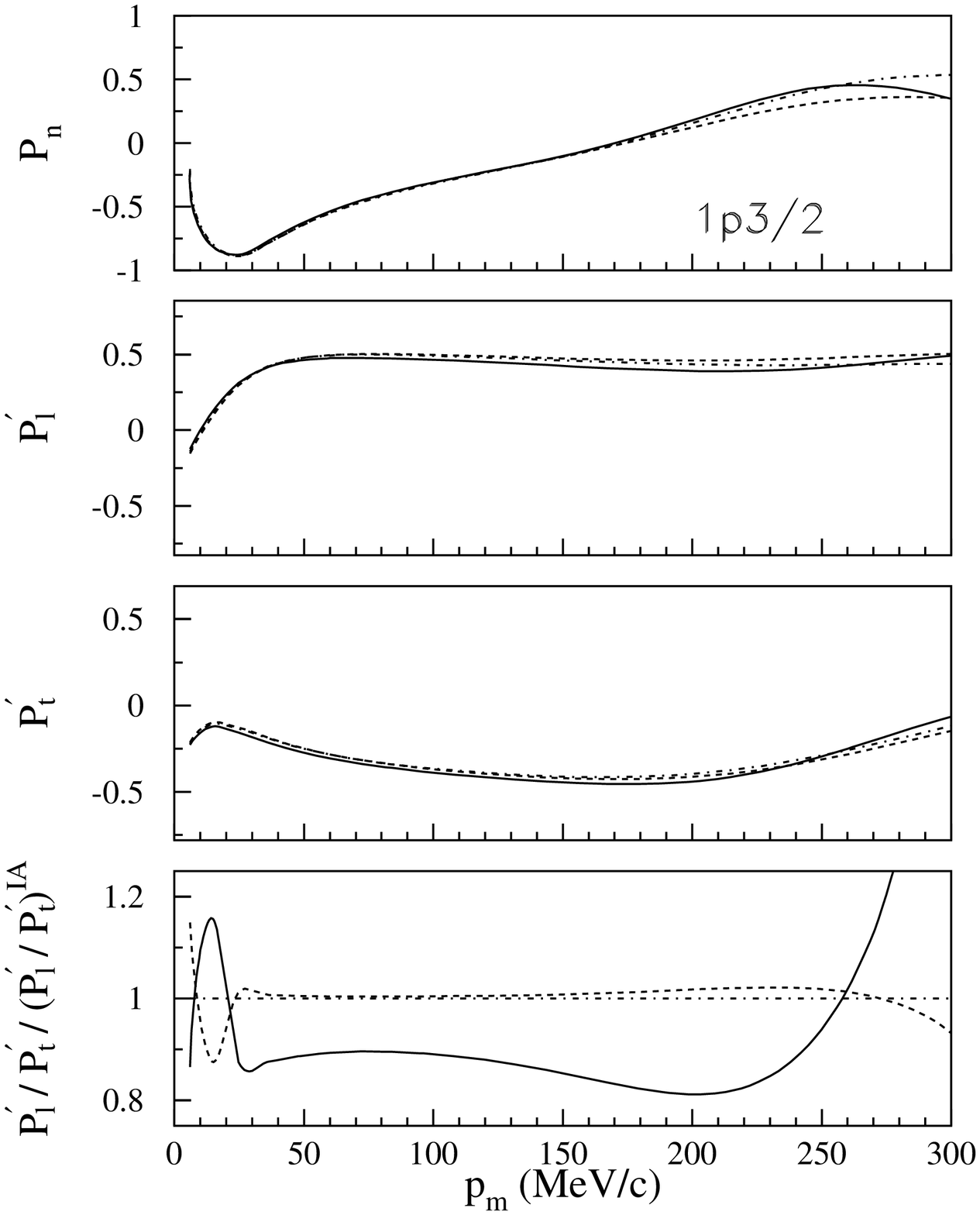}}}
\end{picture}
\end{center}
\caption{The recoil polarization observables for the kinematics of
Figure~\ref{fig:mommainz}. Same line conventions as in Figure~\ref{fig:strmainz}.}
\label{fig:polmainz}
\end{figure}

The question arises whether the relatively strong sensitivity to
two-body currents noticed in Figures~\ref{fig:blom1} and
\ref{fig:blom2} is an intrinsic property of the specific $Q^2$ range
probed or is rather due to the fact that these results were acquired
at smaller values of x and subsequently clearly out of quasi-elastic
conditions.  Results for real quasi-elastic kinematics at about the
same value for the proton's kinetic energy and initial electron energy
are shown in Figures~\ref{fig:mommainz}.  These results are obtained
in quasi-perpendicular kinematics.  Here, the variation in missing
momentum is reached by varying the polar angle $\theta _p$ of the
ejected proton.  For the remainder of this paper
all calculated (reduced) cross sections are shown for full sub-shell
occupancy, i.e. $S_{lj}=(2j+1)$.  It becomes clear that under strict
quasi-elastic conditions the effect of the subnuclear
degrees-of-freedom is substantially smaller than at smaller values of
Bjorken x.  It is worth mentioning that the
results of Figure~\ref{fig:mommainz} are obtained in more transverse
kinematics than those of Figures~\ref{fig:blom1} and
\ref{fig:blom2}.   Accordingly, with the eye on minimizing the effect of
two-body currents it is very essential to move in quasi-elastic
conditions {\sl even if this implies shifting to kinematics which is
conceived to be transverse.}  Another observation is that the extent
to which to the two-body currents are important does not depend on the
polar angle $\theta _N$ of the ejected nucleon.  Figures
\ref{fig:strmainz} and ~\ref{fig:polmainz} show the structure
functions and polarization observables for the same kinematics in
which the cross sections of Figure~\ref{fig:mommainz} are obtained.
In presenting the structure functions we prefer not to divide out any
kinematical variables from the different contributions to the total
angular cross section. Formally, this means that we write
Eq.~(\ref{eq:eep}) in the form
\begin{equation}
{d^5 \sigma \over  d \Omega _N d \epsilon ' d \Omega
_{\epsilon '}} \equiv \sigma_L + \sigma_T + \sigma_{LT} + \sigma_{TT}
+h \left(\sigma ' _{LT} + \sigma ' _{TT} \right)    \; ,
\end{equation}
where the precise definition of all contributing terms is obvious.
This way of presenting results has the outspoken advantage that the
relative contribution of each term in the cross section can be easily evaluated.
Furthermore, possible confusions regarding the adopted definitions for
the different structure functions are avoided.  In line with the
observations made for $^{4}$He and the deuteron \cite{gilad98} case,
the interference terms that are an order of magnitude smaller than
$\sigma _L$ and $\sigma _T$ exhibit the strongest
sensitivity to contributions that go beyond the IA.     
 
\begin{figure}
\begin{center}
\setlength{\unitlength}{1cm}
\begin{picture}(16,12)
\put(-1.00,0.0){\mbox{\epsfysize=15.cm\epsffile{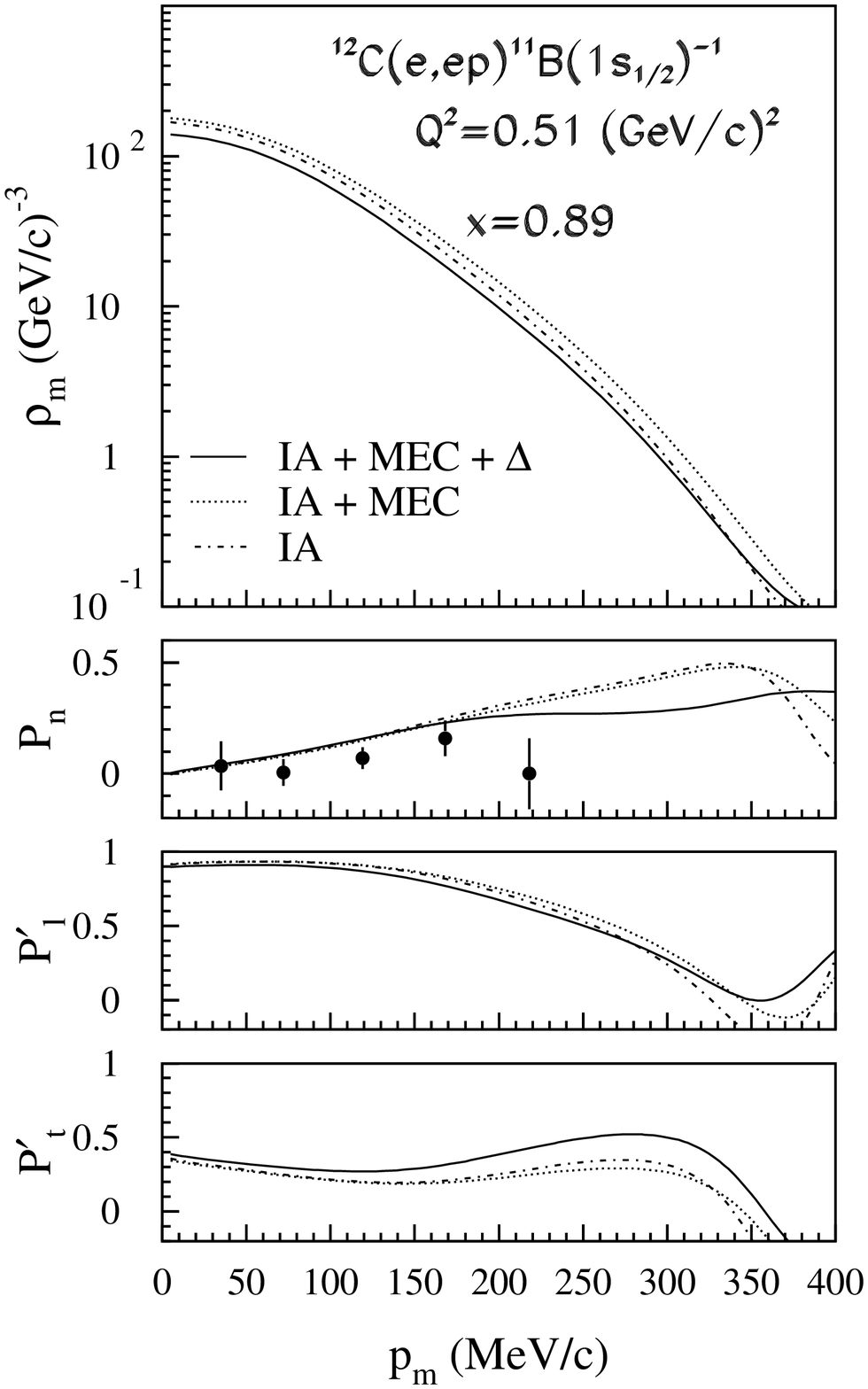}}}
\put(7.50,0.0){\mbox{\epsfysize=15.cm\epsffile{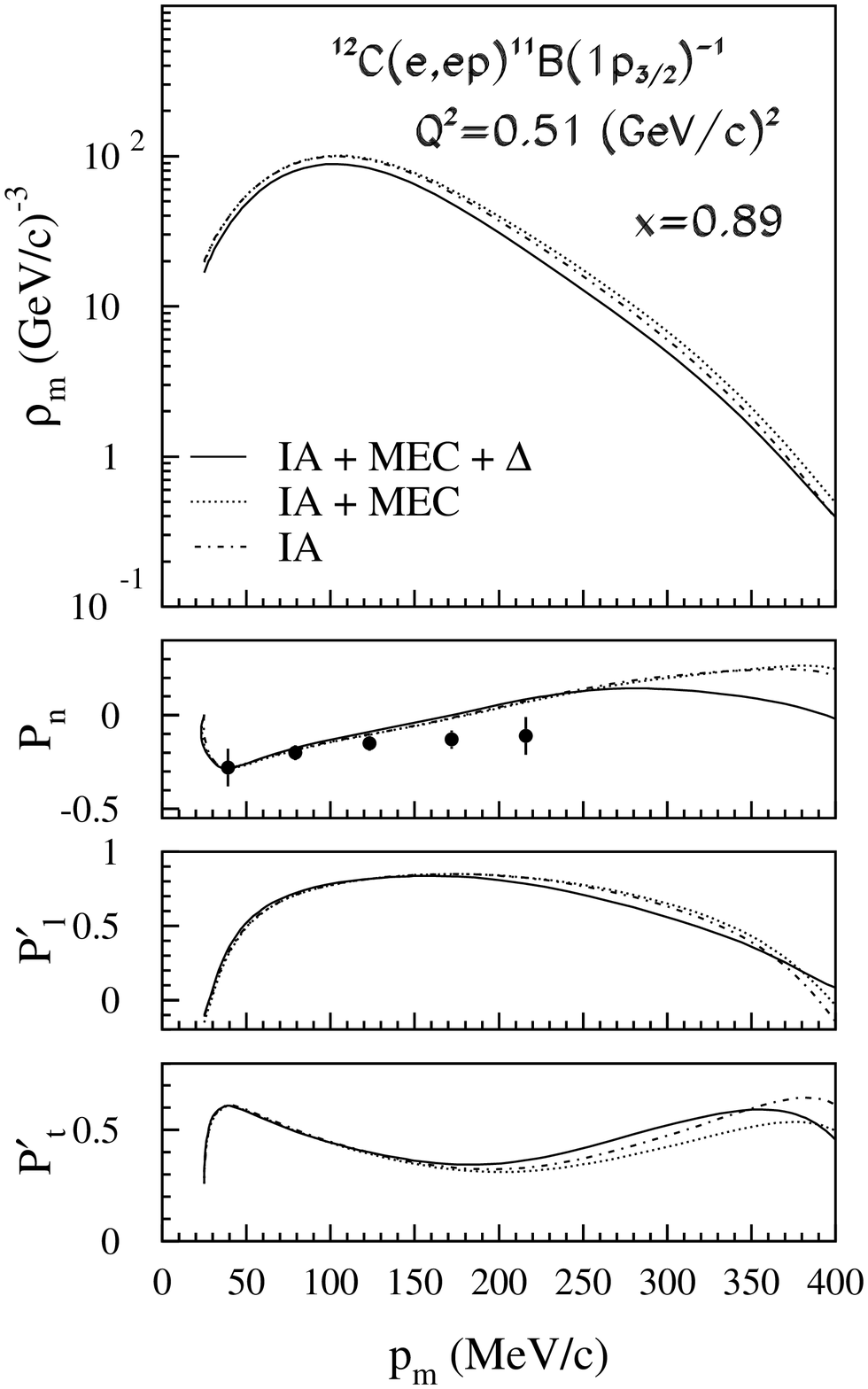}}}
\end{picture}
\end{center}
\caption{Reduced cross section and recoil polarization observables for
$^{12}$C($e,e'p$) in quasi-perpendicular kinematics. The inital
electron energy is $\epsilon$=579~MeV, $\omega$=290~MeV and
q=0.76~GeV/c. The reduced cross sections are normalized for full
1p$_{3/2}$ and 1s$_{1/2}$ occupancy. The data are from Ref.~\protect
\cite{woo98}.}
\label{fig:woo}
\end{figure}

We now turn to the discussion of the recoil polarization observables.
For coplanar kinematics the sole non-vanishing recoil polarization
observables are $P_n,P'_l$ and $P'_t$.  The $P_n$ does vanish
identically in the plane-wave impulse approximation.  As such it allows to
constrain the model assumptions with respect to the final-state
interaction (FSI).  Inspecting Figure~\ref{fig:polmainz} it is clear
that at low missing momenta the qualitative behaviour of the induced
polarization $P_n$ turns out to be relatively insensitive to the
current contributions beyond the impulse approximation.  This result
confirms the findings of the Pavia calculations reported in
Ref.~\cite{mandevil94}.  The first data set for the induced
polarization in finite nuclei has recently become available
\cite{woo98}.  In Figure~\ref{fig:woo} the calculated reduced cross
section and recoil polarization observables are shown for the
kinematics of this experiment.  For knockout from the $1s$ shell,
where the nucleon is mainly located at high nuclear densities, the
influence of the subnuclear d.o.f. is more pronounced than for
$p$-shell knockout. Given that we do not rely on any empirical input
for the description of the FSI, the agreement with the data is
satisfactory but inferior to the quality of agreement that was reached
with the calculations by J.J.~Kelly \cite{woo98}.  These calculations 
constrain the description of the final-state interactions
with the aid of optical potentials derived from the analysis of
hadronic induced reactions. For most observables, the meson-exchange
and isobar effects are acting in opposite directions.  At low missing
momenta the induced polarization exhibits a small sensitivity to the
two-body currents.  At moderate and high missing momenta the $P_n$ is
predicted to exhibit a large sensitivity to two-body currents.  A
similar observation is made for the polarization transfer observables.
Recently it was shown that the recoil proton polarizations for
p($\vec{e},e'\vec{p}$) and d($\vec{e},e'\vec{p}$)n are almost
identical \cite{milbrath98}.  From the theory side, rather small
contributions from the MEC and IC in the double polarization
observables were predicted in the deuteron \cite{gilad98} and $^{4}$He
\cite{laget94} case in quasi-elastic kinematics. Despite the fact that
the polarization transfer variable $P'_l$ and $P'_t$ are related to
the interference structure functions $W'_{TT}$ and $W'_{LT}$ their
predicted sensitivity to the two-body currents looks relatively small.
Recently, the polarization transfer observables raised considerable
interest in that they would provide a direct handle on the (possible)
medium dependence of the nucleon form factors. Indeed, in the
plane-wave impulse approximation (PWIA) it can be shown that for
electroexcitation from a free proton
\begin{equation}
\frac {P_l'}{P_t'} = - \frac {G_M^p} {G_E^p} 
\frac {(\epsilon + \epsilon ') tan \frac {\theta_e} {2}} {2 M_p} \;  .
\end{equation}
It is of the utmost importance to investigate all possible mechanisms
that could bring about changes in the above ratio without being
related to medium modifications of the form factors.  Whereas it was
recently shown that final-state interaction and gauge ambiguities
effects are only marginally affecting the ratio $\frac {P_l'}{P_t'}$
\cite{kelly} the question arises whether meson-exchange and isobaric
currents could bring about any change in the ratio of the double
polarization observables.  Whereas the impact of the subnuclear
d.o.f. on the value of polarization observables looks small in quasi-elastic
kinematics and small missing momenta, the MEC and IC are not necessarily unimportant, especially
for such delicate issues like nucleon form factor studies where the
medium dependency is predicted to be modest \cite{lu98b} and the
projected accuracy of the planned experiments is astounding.  
With the aim of
studying the sensitivity of the ratio $\frac {P_l'}{P_t'}$ to
meson-exchange currents and intermediate isobar creation we have
calculated the following double ratio
\begin{equation}
\frac  { \left( \frac {P_l'}{P_t'} \right)}
       { \left( \frac {P_l'}{P_t'} \right)^{IA}}
\end{equation} 
where ``$IA$'' refers to the calculated ratio in the impulse
approximation, this is when retaining solely the one-body currents in
the calculations but keeping all other ingredients fixed.  As could be
expected, the observed deviation from the IA result enhances as the
missing momentum increases.  The most favorable regime to perform the
recoil polarization measurements for knockout from p-shell orbits, is
probably in the peak of the cross section ($p_m \approx$100~MeV/c).
For the $^{16}$O($e,e'p$)$^{15}$N(1/2$^-$) transition, the deviation
in the ratio $\frac {P_l'}{P_t'}$ that is attributed to the MEC and IC
is about 5-10\% in the peak of the cross section.  For knockout from
its spin-orbit partner ($1p_{3/2}$) the calculated effect is
10-15\%. In both cases, at 200~MeV/c the modifications in the ratio due
to two-body currents has grown to 20\% and with increasing $p_m$ the
effect appears to be rather out of control.

\subsection{$^{16}$O($\vec{e},e'\vec{p}$) results at $Q^2$=0.8~(GeV/c)$^2$ and x$\approx$1} 

\begin{figure}
\begin{center}
\setlength{\unitlength}{1cm}
\begin{picture}(16,12)
\put(-1.00,0.0){\mbox{\epsfysize=9.5cm\epsffile{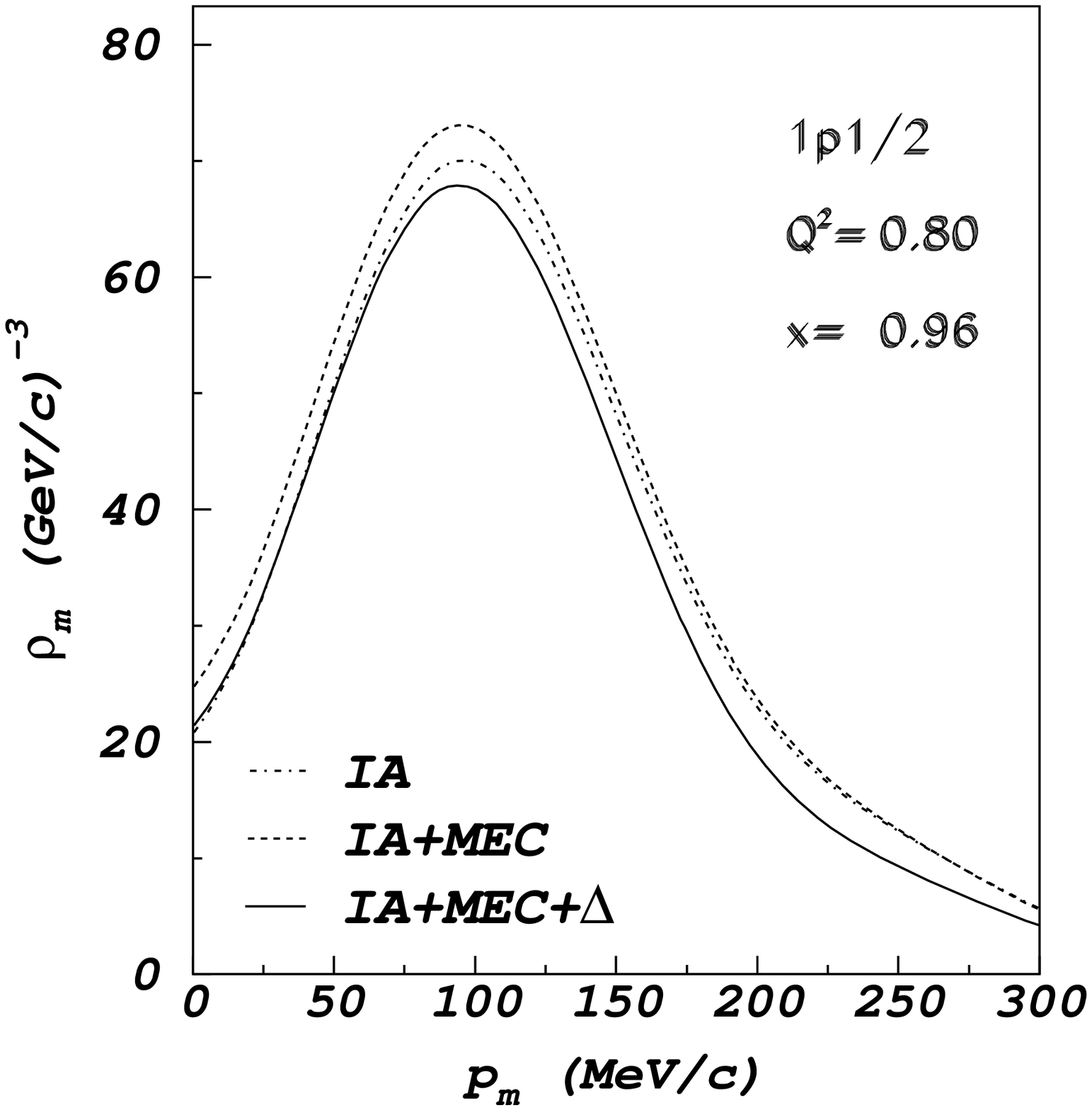}}}
\put(7.50,0.0){\mbox{\epsfysize=9.5cm\epsffile{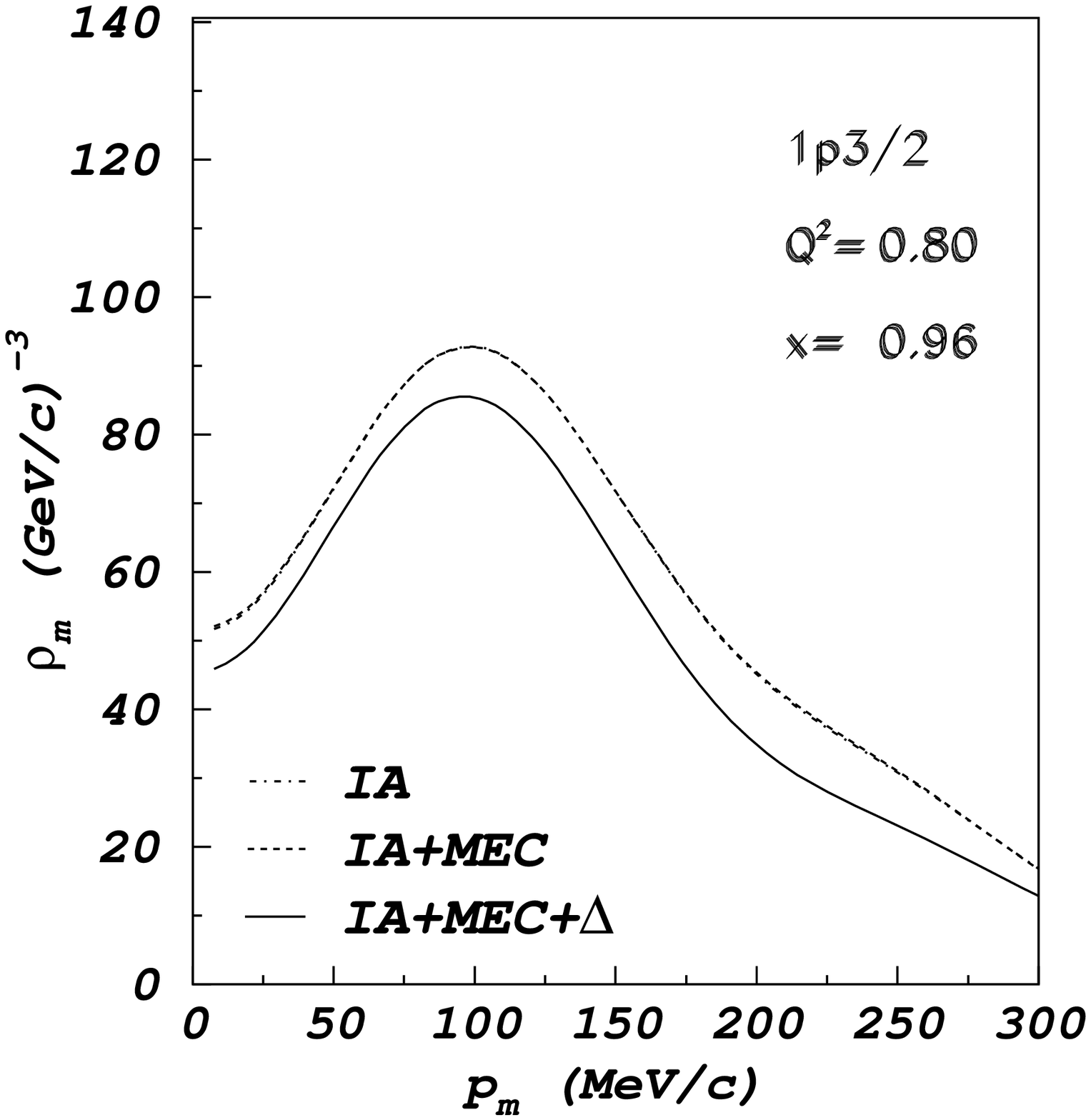}}}
\end{picture}
\end{center}
\caption{Reduced cross sections for $^{16}$O($e,e'p$) from
p-shell states in quasi-perpendicular kinematics.  The electron
kinematics is determined by $\epsilon$=2.445~GeV, $\omega$=445~MeV and
q=1~GeV/c.  The dot-dashed curve shows the result for the impulse
approximation ; in the dashed curve MEC effects are also included, and
the solid curve represents the full calculation
including also IC.}
\label{fig:mome89003}
\end{figure}

\begin{figure}
\centerline{\epsfysize=15.cm \epsfbox{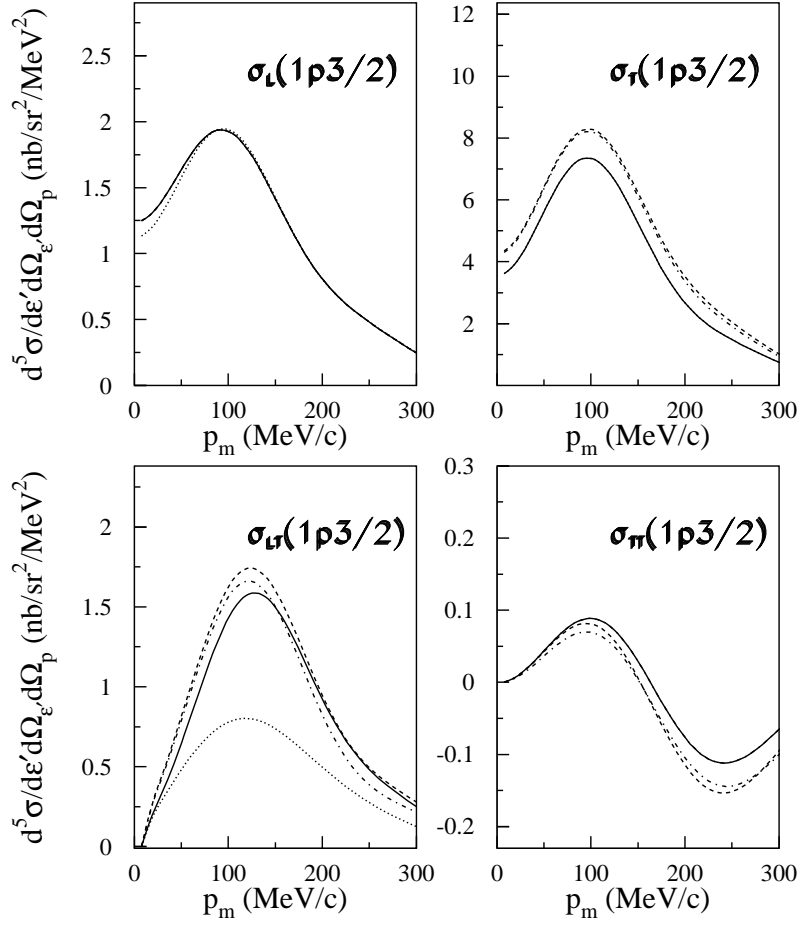}}
\caption{The various structure functions in $^{16}$O($e,e'p$) for
$1p_{3/2}$ knockout in quasi-perpendicular kinematics.  The electron
kinematics is determined by $\epsilon$=2.445~GeV, $\omega$=445~MeV and
q=1~GeV/c.  The dot-dashed curve shows the result for the impulse
approximation ; in the dashed curve MEC effects are also included, and
the solid curve represents the full calculation including also IC. The
dotted line is the result of the full calculation when neglecting the
relativistic corrections in the current operator.}
\label{fig:str2e89003}
\end{figure}

\begin{figure}
\centerline{\epsfysize=15.cm \epsfbox{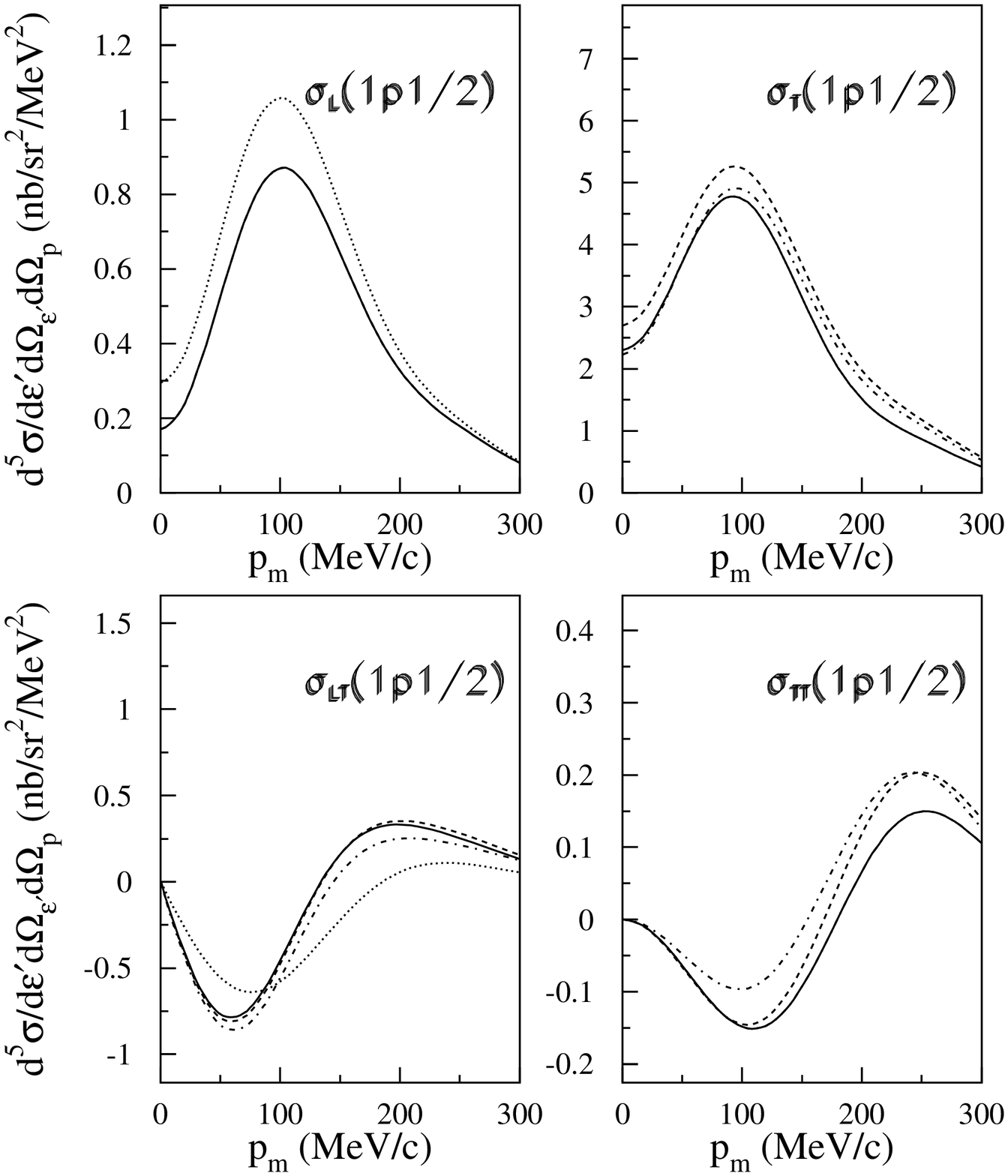}}
\caption{As in Figure~\ref{fig:str2e89003} but now for knockout from
the $1p_{1/2}$ orbital.}
\label{fig:str3e89003}
\end{figure}

\begin{figure}
\centerline{\epsfysize=15.cm \epsfbox{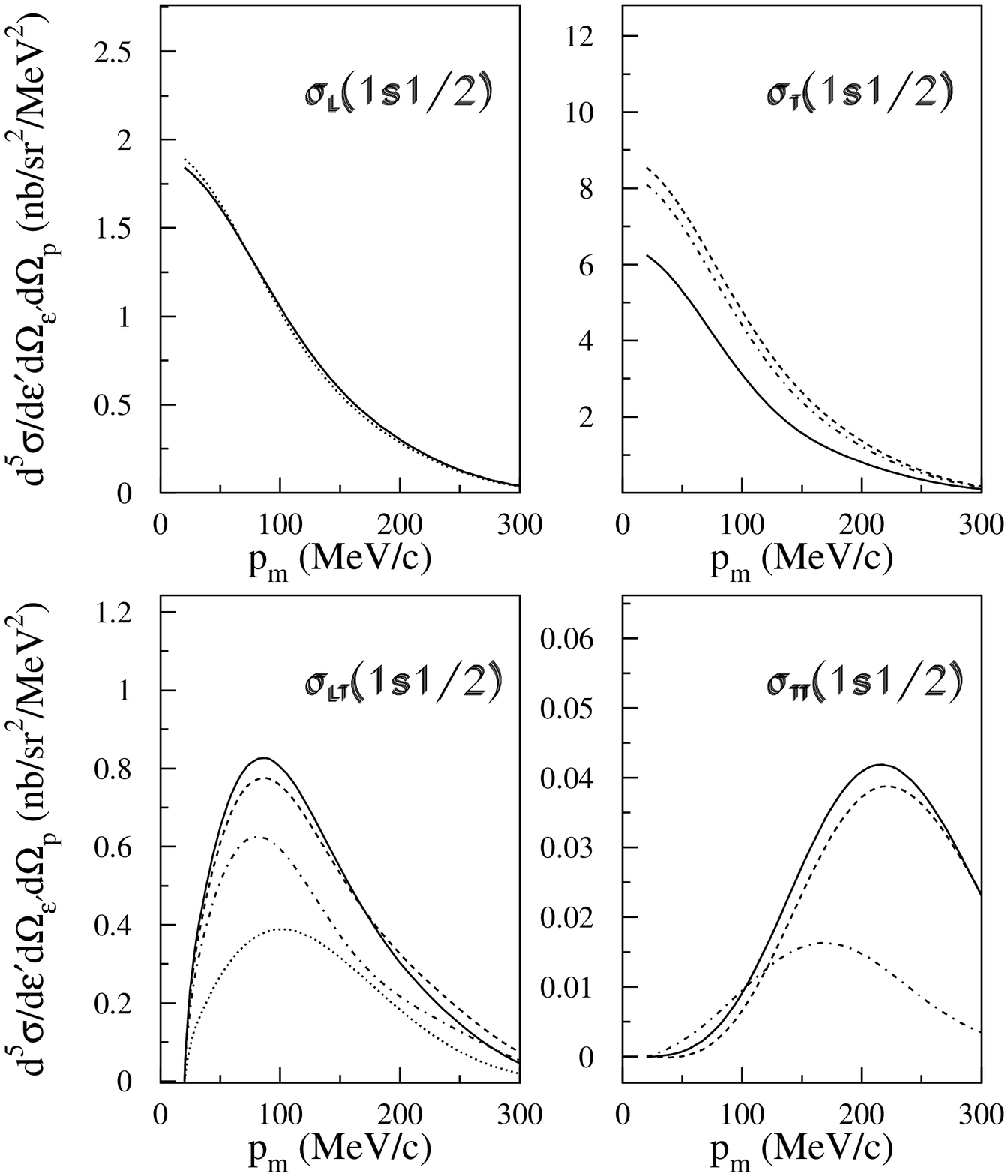}}
\caption{As in Figure~\ref{fig:str2e89003} but now for knockout from
the $1s_{1/2}$ orbital.}
\label{fig:str1e89003}
\end{figure}

\begin{figure}
\setlength{\unitlength}{1cm}
\begin{picture}(16,12)
\put(-1.00,0.0){\mbox{\epsfysize=12.cm\epsffile{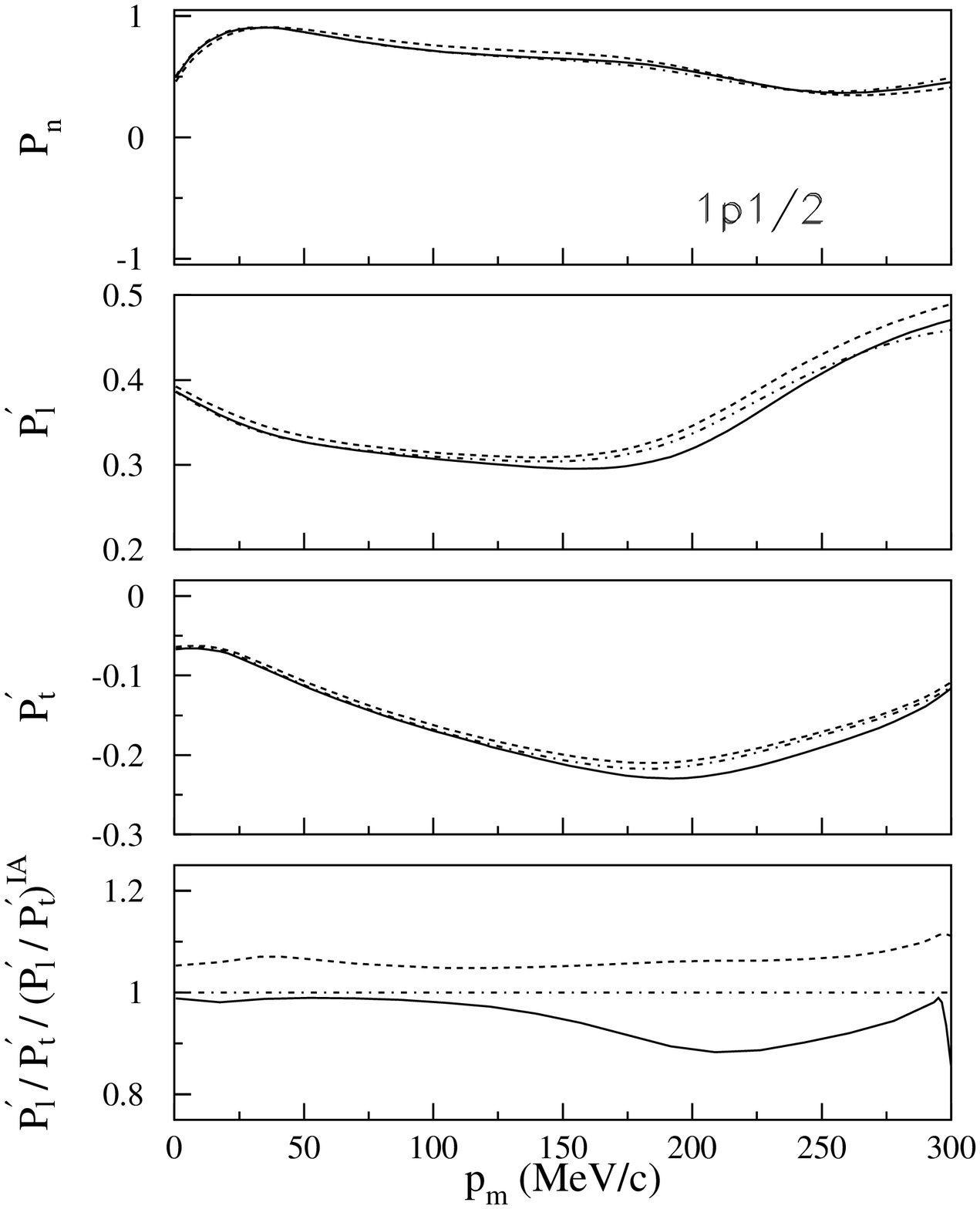}}}
\put(7.50,0.0){\mbox{\epsfysize=12.cm\epsffile{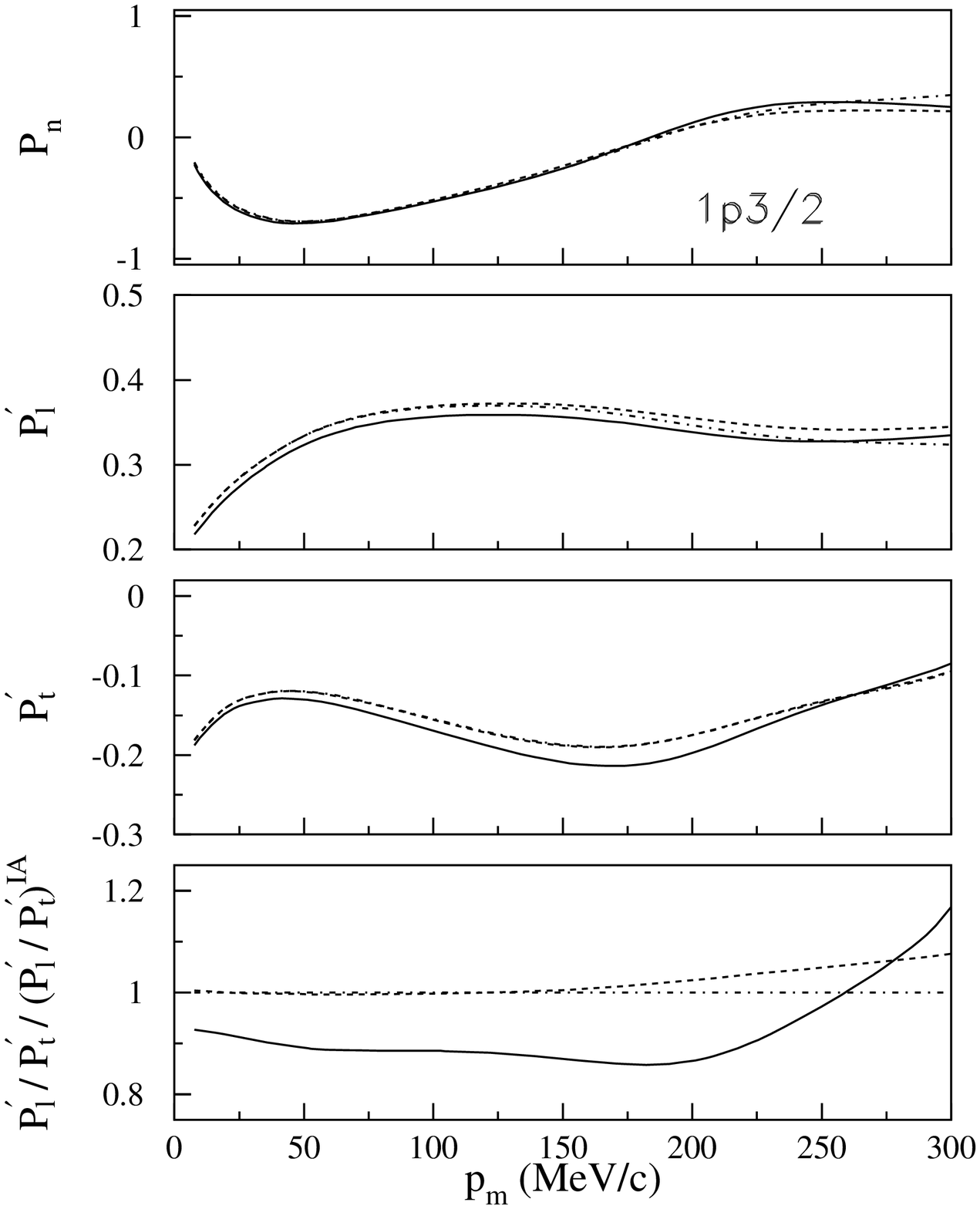}}}
\end{picture}
\caption{Recoil polarization observables as a function of the missing
momentum in quasi-perpendicular
kinematics for knockout from the p-shell orbits in the
$^{16}$O($\vec{e},e'\vec{p}$) reaction at $\epsilon$=2.445~GeV,
$\omega$=445~MeV and q=1~GeV/c. The dot-dashed curve shows the result
for the impulse approximation ; in the dashed curve MEC effects are
also included, and the solid curve represents the full calculation
including also IC.}
\label{fig:pol2a3}
\end{figure}

\begin{figure}
\centerline{\epsfysize=15.cm \epsfbox{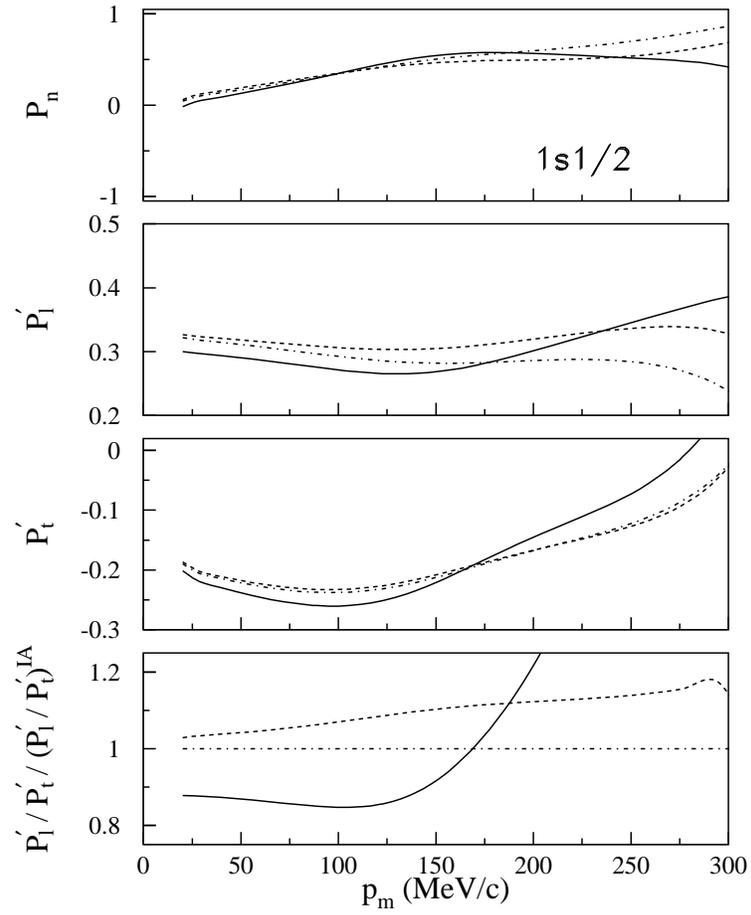}}
\caption{As in Figure~\ref{fig:pol2a3} but for knockout from the
$1s_{1/2}$ orbit.}
\label{fig:sstate}
\end{figure}

We continue our investigations into the role of two-body currents by
considering the kinematics of the TJNAF experiments E89-003 and
E8-9033 \cite{e89033}. The E89-003 experiment has measured the separated structure
functions and momentum distributions for $^{16}$O($e,e'p$) under
quasi-elastic conditions at $\epsilon$=2.445~GeV, $\omega$=445~MeV and
q=1~GeV/c.  In E89-033 on the other hand, the $^{16}$O($\vec{e},e'
\vec{p}$) polarization observables were measured for approximately the
same kinematics.  Figures \ref{fig:mome89003}-\ref{fig:sstate}
summarize the calculated results for the reduced cross sections,
structure functions and recoil polarization observables for knockout
from the two $p$-shell and the $s$-shell state.  In comparing these
results with those obtained in the previous Subsection for
quasi-elastic conditions at lower four-momentum transfer, we can study the
$Q^2$ dependency of the meson and isobar degrees of freedom. In
analogy with the obvervations made at lower $Q^2$, the pion-exchange
currents (``MEC'') increase the cross section and this effect is
completely counterbalanced by the $\Delta$ current that produces the
largest contributions.  The $Q^2$ dependency of the two-body current
effects can maybe be better estimated from the individual structure
functions.  Comparing Figures~\ref{fig:strmainz} and
\ref{fig:str3e89003} one does indeed observe a decreasing trend in the
relative importance of the two-body currents.  A similar tendency is
observed for the recoil polarization observables
(Figure~\ref{fig:pol2a3}.  The deviation in the ratio $\frac
{P_l'}{P_t'}$ in the peak of the cross section is of the order of a
few percent for the ground-state transition (knockout from the
$1p_{1/2}$ orbit), whereas for the excitation of the $3/2^-$ hole
state at 6.32~MeV excitation energy in $^{15}$N (knockout from the
stretched $1p_{3/2}$ orbit) the effect is close to 10\%.

With the aim of studying the medium-dependent effects one could be
tempted to probe the region of highest nuclear density which for a
nucleus like $^{16}$O implies studying the region of $1s_{1/2}$
knockout. In light of the mass-independence of the two-body current
effects that is frequently alluded to, our findings for
$1s_{1/2}$-knockout in $^{16}$O can to a certain extent serve as a
guideline for the effects than could be expected in $^4$He.   From
all orbits studied here, knockout from the $1s_{1/2}$ orbit exhibits
the strongest sensitivity to the two-body currents.  After all, this
observation is not that surprising given its stretched status and the
fact that the region of highest nuclear density is probed.  For
example, the $LT$ and $TT$ interference responses in
Figure~\ref{fig:str1e89003} grow substantially after including the
two-body currents. Indications for the strong sensitivity of the
${LT}$ structure function to two-body currents was reported for the
$^{4}$He case in Ref.~\cite{epstein93}.  Also for the recoil
polarization observables in the missing-energy range of $1s_{1/2}$
knockout (Figure~\ref{fig:sstate}), sizeable contributions from the
meson-exchange and isobar currents are predicted.  For the ratio
$\frac {P_l'}{P_t'}$ the two-body currents bring about a reduction
which is slightly bigger than 10\%.  This is qualitatively very
similar with the result obtained for the ``stretched'' p-shell state
($1p_{3/2}$) (Figure~\ref{fig:pol2a3}) so that some general
qualitative behaviour for the sensitivity of the $\frac {P_l'}{P_t'}$
ratio to multi-nucleon currents seems to emerge.  Indeed, at low
missing momenta and quasi-elastic conditions the two-body currents
reduce the ratio $\frac {P_l'}{P_t'}$, an effect that slowly decreases
as one goes to higher $Q^2$ and is larger for knockout from
``stretched'' than from the ``jack-knifed'' single-particle orbits.
Finally we turn our attention to the role of the relativistic
corrections (Section~\ref{sec:current}) in the one-body current
operator.  The dotted lines in Figures~\ref{fig:str2e89003},
\ref{fig:str3e89003} and \ref{fig:str1e89003} show the calculated
contributions to the cross section when neglecting the relativistic
corrections in the one-body current operator, i.e. when adopting
$C_1(q,Q^2)=C_3(q,Q^2)=1$ and $C_2(q,Q^2)=0$ in
Eq.~(\ref{eq:redcurrent}).  When comparing these curves with the solid
ones one can estimate the effect of the relativistic corrections in
the one-body current.  For the transverse responses $\sigma _T$ and
$\sigma _{TT}$ the Foldy-Wouthuysen prescription does not lead to any
relativistic correction into lowest order.  Whereas the spin-orbit term in the
charge-density operator produces rather small corrections in the
longitudinal response $\sigma _L$ in the considered kinematics, the
longitudinal-transverse response $\sigma _{LT}$ for the ``stretched''
single-particle states $1p_{3/2}$ and $1s_{1/2}$ is doubled after
including the relativistic corrections in the charge-density operator.
A strong sensitivity of the $LT$ response to relativistic corrections
was earlier found for $d(e,e'p)$ \cite{gilad98,hummel94,ducret94}.


\subsection{Neutron knockout}
\begin{figure}
\begin{center}
\setlength{\unitlength}{1cm}
\begin{picture}(16,12)
\put(-1.00,0.0){\mbox{\epsfysize=11.cm\epsffile{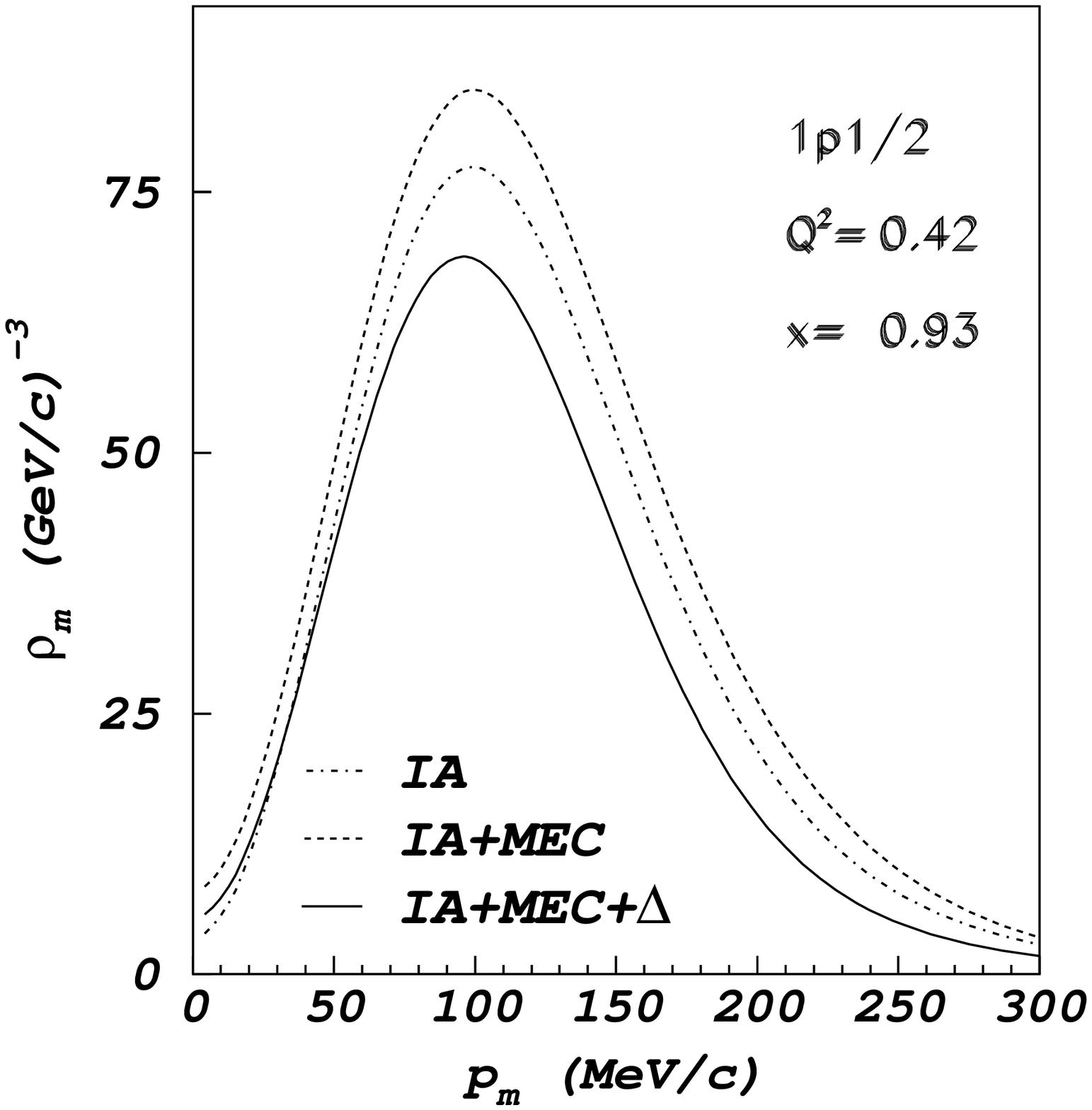}}}
\put(7.50,0.0){\mbox{\epsfysize=11.cm\epsffile{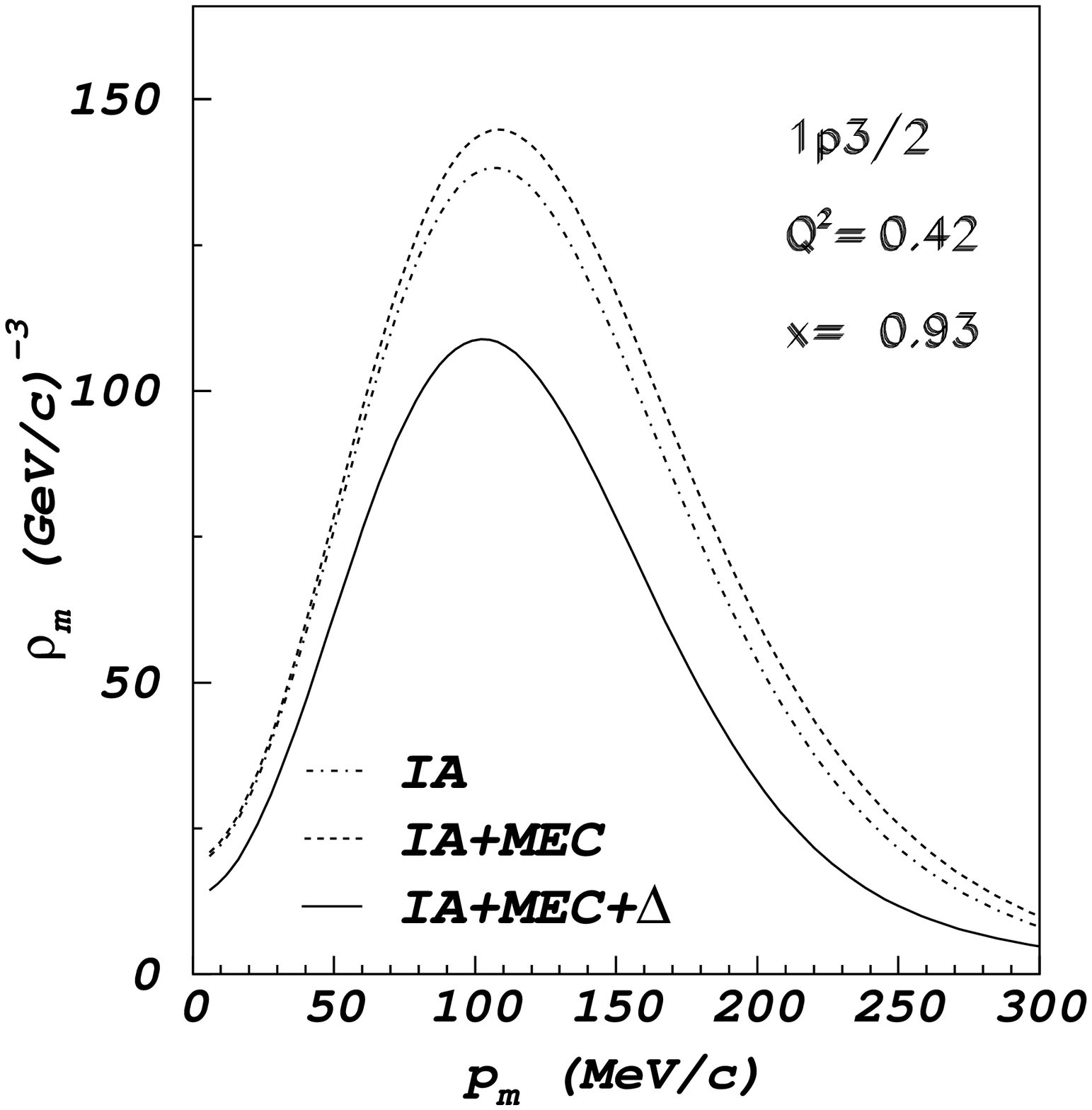}}}
\end{picture}
\end{center}
\caption{Reduced cross sections for $^{16}$O($e,e'n$) from
p-shell states in quasi-perpendicular kinematics.  The electron
kinematics is determined by $\epsilon$=0.855~GeV, $\omega$=240~MeV and
q=0.69~GeV/c.  The dot-dashed curve shows the result for the impulse
approximation ; in the dashed curve MEC effects are also included, and
the solid curve represents the full calculation
including also IC. The curves are normalized for full sub-shell occupancy.}
\label{fig:een}
\end{figure}

To our knowledge, beyond the giant resonance region no ($e,e'n$)
measurements have been made.  Nevertheless, neutron knockout
investigations with moderate energy resolution could be done at MAMI,
BATES or TJNAF.  Results for neutron knockout in kinematic conditions
that can be reached with electron beam energies below 1~GeV are shown
in Figure~\ref{fig:een}.  These results are obtained with a
neutron formfactor $G_E^n$=0.  The $(e,e'n)$ calculations were done in
the same kinematical conditions for the ($e,e'p$) curves of
Figure~\ref{fig:mommainz}.  The reduced cross sections for neutron
knockout are subject to larger two-body current corrections than
corresponding proton results.  As a matter of fact, the absolute
magnitude of the two-body current contributions is comparable for
proton and neutron knockout. 
The fact that $\left( G_M^p/ G_M^n \right)^2 \approx$2 and the
absence of a substantial neutron charge, however, makes them to be relatively
more important in the neutron knockout channel.  As could be expected
on the basis of the isospin structure of the underlying current
operators, the qualitative behaviour is similar for proton and neutron
knockout.  The net reduction of the cross sections displayed in
Figure~\ref{fig:een} is of the order 10-20\%.  Note that
coupled-channel-effects between the proton knockout channels and the
considerably weaker ($e,e'n$) channels are expected to compensate
largely for the loss of strength brought about by the meson and isobar
degrees-of-freedom \cite{kelly2,jan89}.  Our ($e,e'n$) results are
qualitatively similar to those reported in \cite{boffi92}. The MEC are
observed to increase the cross section, an effect that is largely
overcompensated by the isobaric current that works in the opposite
direction.  This reduction can be attributed to the fact that the
one-body magnetization current and the isobaric current interfere
destructively.

\subsection{$^{208}$Pb($e,e'p$) results at $Q^2$=0.3~(GeV/c)$^2$ and x$\approx$1} 
\begin{figure}
\begin{center}
\setlength{\unitlength}{1cm}
\begin{picture}(16,12)
\put(-1.00,0.0){\mbox{\epsfysize=11.5cm\epsffile{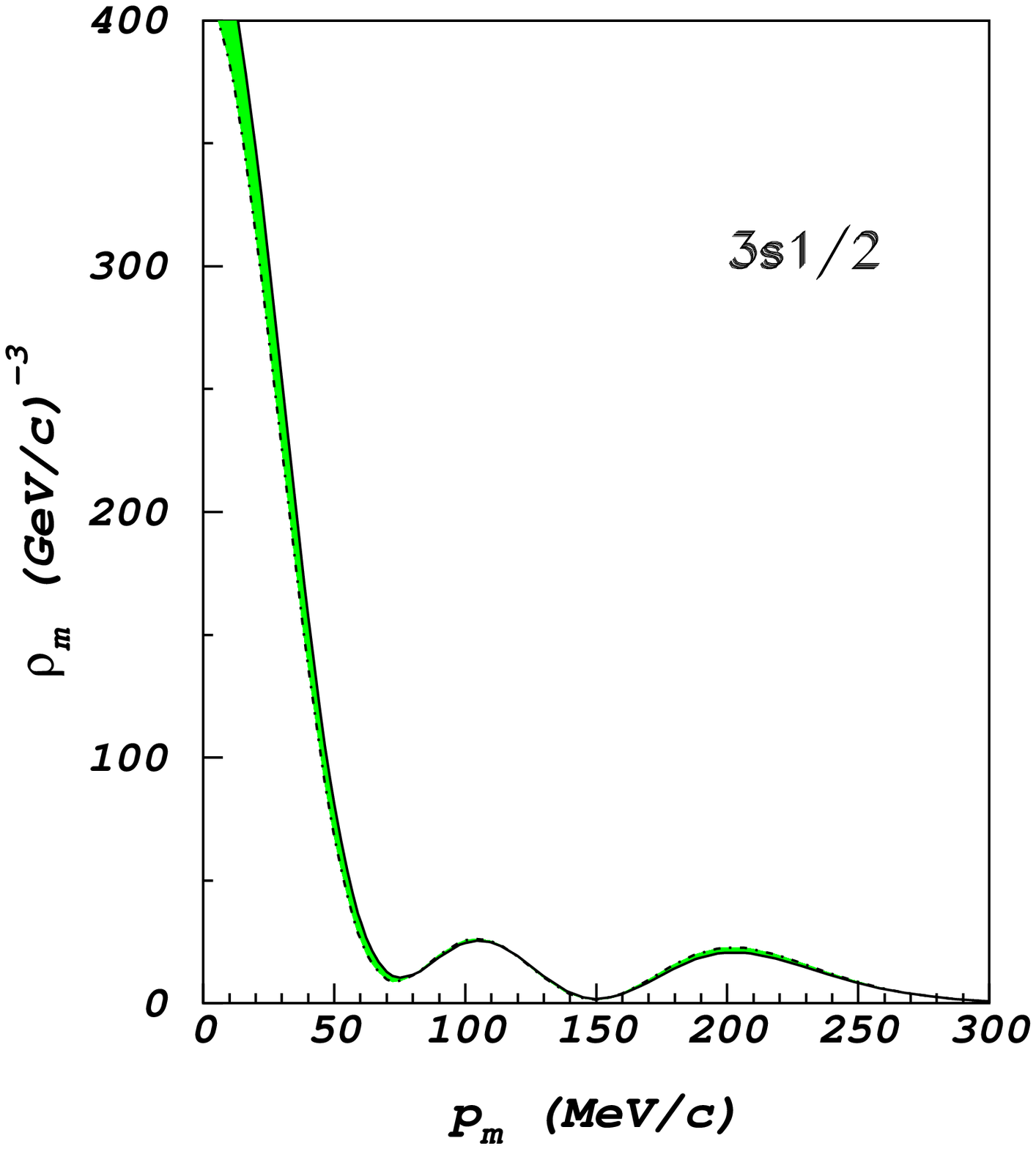}}}
\put(7.50,0.0){\mbox{\epsfysize=11.5cm\epsffile{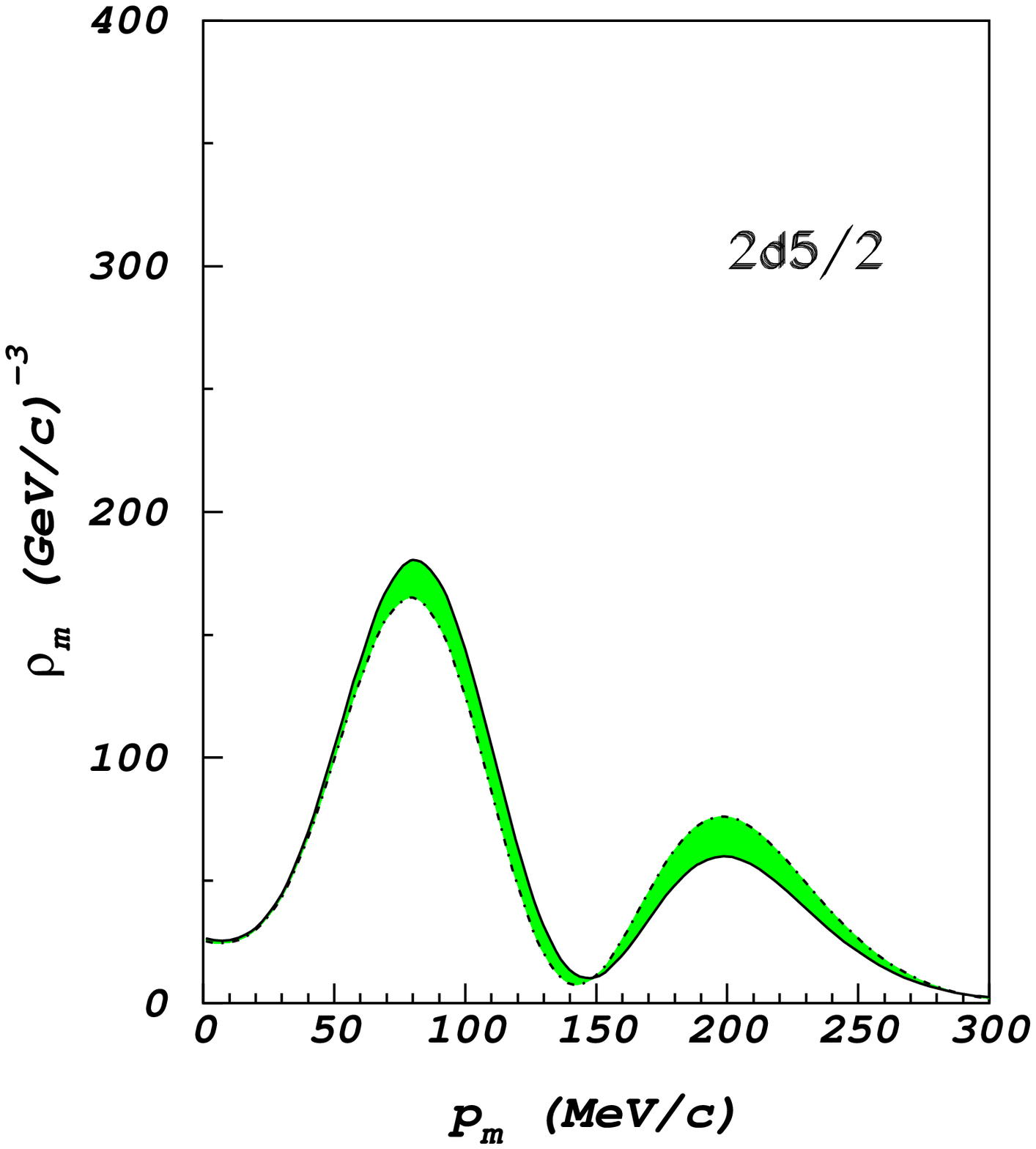}}}
\end{picture}
\end{center}
\caption{Reduced cross sections for $^{208}$Pb($e,e'p$) from
in quasi-perpendicular kinematics.  The electron
kinematics is determined by $\epsilon$=0.462~GeV, $\omega$=170~MeV and
q=0.57~GeV/c.  The dot-dashed curve shows the result for the impulse
approximation, whereas the solid curve is the result of a calculation
that includes also MEC and IC. The curves are normalized for full
sub-shell occupancy.}
\label{fig:pb208}
\end{figure}

In this section we report on calculations for an $^{208}$Pb($e,e'p$)
experiment that was done at NIKHEF.  This experiment was performed in
very transverse kinematics ($\theta_e=$96$^o$) at an incident electron
energy of 462.14~MeV and near quasi-elastic conditions
($T_p$=161.~MeV, x=0.94).  As this corresponds with very
``transverse'' kinematics we consider this as a 
worst-case-scenario benchmark for investigating the sensitivity to
two-currents that could be expected in quasi-elastic ($e,e'p$) from
heavy nuclei. This is particularly important in view of the fact that
the $^{208}$Pb target is sometimes advocated as an appropriate
surrogate for nuclear matter studies.  In comparison with studies on
light nuclei, for heavy nuclei there is a price to pay in that Coulomb
distortion effects and reduced nucleon transparencies make both the
electromagnetic and the FSI part of the reaction process more
difficult to handle and subject to likely enhanced theoretical
uncertainties.  Here, we only want to study the {\em relative
contribution} from the two-body currents that could be expected for a
heavy nucleus like $^{208}$Pb.  For computational reasons we have
neglected the Coulomb distortion effects.  Indeed, an exact treatment
of these would imply that the two-body matrix elements of
Eq.~(\ref{eq:mattwo}) that involve a sum over all nucleons in the
target nucleus, are to be calculated for a whole range of q values.
Whereas calculations of this type could be performed at lower momentum
transfer \cite{veerle2}, the amount of multipoles required at higher
momentum transfer makes it an enormous computational task.
Figure~\ref{fig:pb208} displays the reduced $^{208}$Pb($e,e'p$) cross
section for knockout from the $3s_{1/2}$ and $2d_{5/2}$ orbits. It
becomes clear that the predicted two-nucleon effects are modest and of the same
relative size in comparison with the findings for light nuclei.
Whereas the transparency for nucleon knockout has been established to
decrease as the mass number increases \cite{potterfeld} the 
impact of the two-body currents on single-nucleon knockout appears to
be rather mass independent.  

\section{Summary and Conclusions}
\label{sec:conclusions}
Within the context of ongoing research activities in two-nucleon and
pion production from nuclei a fair understanding of the two-nucleon
currents in finite nuclei has been reached.  With this knowledge the
contribution of two-body currents to exclusive ($e,e'p$) and ($e,e'n$)
reactions can be calculated with some confidence.  Reasonable
estimates of their contribution is essential in view of the fact that
subnuclear d.o.f. frequently represent unwanted background that could
be at the origin of ambiguities in the interpretation of
($\vec{e},e'\vec{p}$) studies. Most of our investigations were carried
out in quasi-elastic kinematics and momentum transfers $\mid \vec{q}
\mid \le$~1~GeV/c. The effect of two-body currents dramatically
increases as one moves out of quasi-elastic conditions and/or higher
missing momenta are probed. In quasi-elastic kinematics and the whole
range of $Q^2$ values studied here, the two-body currents 
decrease the $(e,e'p)$ differential cross sections when lower missing
momenta are probed.  As the shape does not seem to be substantially
affected by the two-body currents it appears virtually impossible to
acquire experimental evidence for the role of two-body current effects
from plain $(e,e'p)$ cross-section measurements in quasi-elastic
kinematics.  At the same time, our findings imply that the
spectroscopic factors as they are usually derived from IA calculations
are subject to corrections that would make them bigger.
These corrections are not unimportant and exhibit a $Q^2$ dependence.
Even at four-momentum transfers of $Q^2 \approx$~0.8~(GeV/c)$^2$ the
effect of the two-body currents on the spectroscopic factor can be as
large as 20\%.  The role of the two-body currents can be made more
explicit by measuring the interference response functions that,
admittedly, represent a rather small part in the total angular cross
section.
The induced polarization $P_n$ exhibits a moderate sensitivity to the
two-body currents at low missing momenta $p_m \le 200$~MeV/c.
Therefore, it retains its status as a proper variable to constrain the
final-state interaction effects.  Similar sort of sensitivities are
found for the polarization transfer variables.  In the light of
exploiting these variables to increase our understanding of the
(possible) medium
dependency in the nucleon form factors, the effect of the meson and
isobar degrees of freedom on the ratio $\frac {P_l'}{P_t'}$ can be
boiled down to a few percent at higher values of $Q^2$.  Under typical
MAMI kinematics, which implies four-momentum transfers of the order
$Q^2 \le $0.50~(GeV/c)$^2$, the impact of subnuclear degrees of
freedom on the $\frac {P_l'}{P_t'}$ ratio is predicted to be of the
order of 10\%.  Of all pion-related two-body currents studied here the
$\Delta$-isobar current has by far the strongest impact on the
($\vec{e},e'\vec{p}$) and ($e,e'n$) observables.  Multi-coincidence
experiments of the type ($\gamma^{(*)},NN$) and ($\gamma^{(*)},N\pi$)
at higher four-momentum transfers $Q^2$ would greatly help in further
constraining the model assumptions with respect to the $\Delta$-isobar
current operator when reaching higher energy and momentum transfers.

{\bf Acknowledgement}

This work was supported by the Fund for Scientific Research of
Flanders under Contract No 4.0061.99 and the University Research
Council.

\begin{thebibliography}{99}
\bibitem{kelly96} J.J. Kelly, Adv. Nucl. Phys. {\bf 23} (1996) 75.
\bibitem{vijay98}  
V. Phandaripande, I. Sick and P.K.A. de Witt Huberts,
Rev. Mod. Phys. {\bf 69} (1997) 981.
\bibitem{lu98a} D.H.~Lu {\em et al.}, Nucl. Phys. {\bf A634} (1998)
443.
\bibitem{lu98b} D.H.~Lu {\em et al.}, Phys. Lett. {\bf B417} (1998)
217.
\bibitem{laget94}
J.M. Laget, Nucl. Phys. {\bf A579} (1994) 333.
\bibitem{e89033} TJNAF experiment E89-003 ``Measurement of Recoil
Polarization in the $^{16}$O($e,e'p$) Reaction with 2.4 GeV
Electrons'' (spokesperson C.~Glashausser).
\bibitem{e93049} TJNAF experiment E93-043 ``Polarization Transfer in
the Reaction $^{4}$He($e,e'p$)$^3$H in the Quasi-elastic Scattering
Region'' (spokesperson J.~Van~Den~Brand)
\bibitem{e91006} TJNAF experiment E91-006 ``Study of Nuclear Medium
Effects by Recoil Polarization up to High Momentum Transfers''
(spokesperson A.~Saha)
\bibitem{kelly1} J.J. Kelly, Phys. Rev. C {\bf 56} (1997) 2672.
\bibitem{kelly2} J.J. Kelly, nucl-th/980990
\bibitem{suzuki89} Toshio Suzuki, Nucl. Phys. {\bf A495} (1989) 581.
\bibitem{boffi90}
S. Boffi, C. Giusti, F.D. Pacati and M. Radici, Nucl. Phys. {\bf A518}
(1990) 639.
\bibitem{boffi90a}
S. Boffi and M. Radici, Phys. Lett. {\bf B242} (1990) 151.
\bibitem{boffi92}
S. Boffi, M. Radici, J.J. Kelly and T.M. Payerle, Nucl. Phys. {\bf
A539} (1992) 597.
\bibitem{epstein93} M.B. Epstein {\em et al.}, Phys. Rev. Lett. {\bf
70} (1993) 2868.
\bibitem{veerle1} V. Van der Sluys, J. Ryckebusch and M. Waroquier,
Phys. Rev. C {\bf 54} (1996) 1322.
\bibitem{veerle2} V. Van der Sluys, J. Ryckebusch and M. Waroquier,
Phys. Rev. C {\bf 49} (1994) 2695.
\bibitem{raskin} A.S. Raskin and T.W. Donnelly, Ann. of Phys. {\bf
191} (1989) 78.
\bibitem{pickle87} A. Picklesimer and J.W. Van Orden, Phys. Rev. C
{\bf 35} (1987) 266.
\bibitem{boffigiusti} S. Boffi, C. Giusti, F.D. Pacati and M. Radici,
Electromagnetic Response of Atomic Nuclei, Oxford Studies in Nuclear
Physics (Clarendon Press, Oxford, 1996).
\bibitem{potterfeld} 
D. Potterfeld {\em et al.}, in {\em Proceedings of
the Conference on Perspectives in Hadronic Physics}, edited by
S. Boffi, C. Ciofi degli Atti and M. Giannini, (World Scientific,
Singapore, 1998), 164.
\bibitem{Ryc88} J. Ryckebusch, M. Waroquier, K. Heyde, J. Moreau and
D. Ryckbosch, Nucl. Phys. {\bf A476} (1988) 237.
\bibitem{Mah} C. Mahaux and H. Weidenm\"{u}ller, {\em in} Shell model
approach to nuclear reactions (North-Holland, Amsterdam, 1969).
\bibitem{amaro98} J.E. Amaro, M.B. Barbaro, J.A. Caballero,
T.W. Donnelly and M. Molinari, Nucl. Phys. {\bf A643} (1998) 349. 
\bibitem{fearing94}
H.W. Fearing, G.I. Poulis and S. Scherer, Nucl. Phys. {\bf A570}
(1994) 657.
\bibitem{boffi98} S. Boffi, F. Capuzzi, P. Demetriou and M. Radici,
Nucl. Phys. {\bf A637} (1998) 585.
\bibitem{jeschonn98} S. Jeschonnek and T.W. Donnelly, Phys. Rev. C
{\bf 57} (1998) 2483.
\bibitem{wright} 
Yanhe Jin and D.S. Onley, Phys. Rev. C {\bf 50} (1994) 377.
\bibitem{udias}
J.M. Ud\'{i}as, P. Sarriguren, E.~Moya~de~Guerra, E.~Garrido and
J.A. Caballero, Phys. Rev. C {\bf 48} (1993) 2731.
\bibitem{johans}
M.~Hedayati-Poor, J.I. Johansson and H.S.~Sheriff, Phys. Rev. C {\bf
51} (1995) 2044.
\bibitem{mcdermott} J.P.~McDermott, Phys. Rev. Lett. {\bf 65} (1990) 1991.
\bibitem{roccojoe} J. Carlson and R. Schiavilla, Rev. Mod. Phys. {\bf
70} (1998) 743.
\bibitem{jan97} J. Ryckebusch, V. Van der Sluys, K. Heyde, H. Holvoet,
W. Van Nespen, M. Waroquier and M. Vanderhaeghen, Nucl. Phys. {\bf
A624} (1997) 581.
\bibitem{marcucci98} L.E. Marcucci, D.O. Riska and R. Schiavilla,
Phys. Rev. C {\bf 58} (1998) 3069.
\bibitem{stoler93} P.~Stoler, Phys. Rep. {\bf 226} (1993) 103.
\bibitem{frolov99} V.V.~Frolov {\em et al.}, Phys. Rev. Lett. {\bf 82}
(1999) 45.
\bibitem{Oset}
E. Oset, H. Toki and W. Weise, Phys. Rep. {\bf 83} (1982) 281.
\bibitem{dekker} 
M.J. Dekker, P.J. Brussaard and J.A. Tjon, Phys. Rev.
C {\bf 49} (1994) 2650.
\bibitem{thomas} 
T. Wilbois, P. Wilhelm, and H. Arenh\"ovel, Phys. Rev. C {\bf 54}
(1996) 3311. 
\bibitem{reply}
J. Ryckebusch, L. Machenil, M. Vanderhaeghen, V. Van der Sluys, and M.
Waroquier, Phys. Rev. C {\bf 54} (1996) 3313. 
\bibitem{osterfeld}
B. K\"{o}rfgen, P. Oltmanns, F. Osterfeld and T. Udagawa, Phys. Rev. C
{\bf 55} (1997) 1819.
Rev. C {\bf 50} (1994) 1637.
\bibitem{bianchi}
N. Bianchi {\em et al.}, Phys. Rev. C {\bf 54} (1996) 1688.
\bibitem{koch}
J.K. Koch and N. Ohtsuka, Nucl. Phys. {\bf A435} (1985) 765 and J.
Koch {\em in} Moderns Topics in Electron Scattering, Eds. B. Frois and
I. Sick, (World Scientific, Singapore) (1991) 28.
\bibitem{oconnell}
J.S. O'Connell {\em et al.}, Phys. Rev. C {\bf 35} (1987) 1063.
\bibitem{cenni84} R. Cenni, P. Christillin and G. Dillon,
Phys. Lett. {\bf B139} (1984) 341.
\bibitem{korfgen94}
B. K\"{o}rfgen, F. Osterfeld and T. Udagawa, Phys. Rev. C {\bf 50}
(1994) 1637.
\bibitem{chen}
C.R. Chen and T.-S.H. Lee, Phys. Rev. C {\bf 38} (1988) 2187.
\bibitem{kondra94} L.A. Kondratyuk, M.I. Krivoruchenko, N. Bianchi,
E. De Sanctis and V. Muccifora, Nucl. Phys. {\bf A579} (1994) 453.
\bibitem{oset87} 
E. Oset and L.L. Salcedo, Nucl. Phys. {\bf A468} (1987) 631.
\bibitem{douglas98}
I.J.D. MacGregor {\em et al.}, Phys. Rev. Lett. {\bf 80} (1998) 245.
\bibitem{jana568}
J. Ryckebusch, M. Vanderhaeghen, L. Machenil and M. Waroquier,
Nucl. Phys. {\bf A568} (1994) 828-854. 
\bibitem{guenther} G. Rosner, {\em in} Proc. Conf. on Perspectives in Hadronic Physics,
ICTP Trieste, Italy, May 12-16, 1997, eds. S. Boffi, C. Ciofi degli
Atti and M.M. Gianninni (World Scientific, Singapore) 185.
P. Bartsch {\em et al.}, {\em ``Investigation of short-range 
nucleon-nucleon correlations using the reaction
$^{16}$O(e,e$'$pp)$^{14}$C in super-parallel kinematics'' } MAMI
proposal A1/1-97 (spokesperson G. Rosner), 1997.  
\bibitem{kaschl71} G. Kaschl, G. Mairle, H. Mackh, D. Hartwig and
U. Schwinn, Nucl. Phys. {\bf A178} (1971) 275.
\bibitem{gercoprl2} 
C.J.G. Onderwater {\em et al.}, Phys. Rev. Lett. {\bf
81} (1998) 2213.
\bibitem{gearhart} C.C.Gearhart, PhD thesis, Washington University
(St. Louis, 1994), unpublished and W. Dickhoff, private communication.
\bibitem{blom98} 
K.I. Blomqvist {\em et al.}, Phys. Lett. {\bf B421} 
(1998) 71.
\bibitem{blomqvist95} K.I. Blomqvist {\em et al.}, Phys. Lett. {\bf
B344} (1995) 85.
\bibitem{forest} T. de Forest, Nucl. Phys. {\bf A392} (1983) 232.
\bibitem{kester96}
L.J.H.M. Kester {\em et al.}, Phys. Lett. {\bf B366} (1996) 44.
\bibitem{garino92} G. Garino {\em et al.}, Phys. Rev. C {\bf 45}
(1992) 780.
\bibitem{leuschne94} M. Leuschner {\em et al.}, Phys. Rev. C {\bf 49}
(1994) 955.
\bibitem{caba98} J.A. Caballero, T.W. Donnelly, E. Moya de Guerra and
J.M. Ud\'{i}as, Nucl. Phys. {\bf A643} (1998) 189.
\bibitem{gilad98} S. Gilad, W. Bertozzi and Z.-L. Zhou,
Nucl. Phys. {\bf A631} (1998) 276c.
\bibitem{mandevil94} J. Mandeville {\em et al.}, Phys. Rev. Lett. {\bf
72} (1994) 3325.
\bibitem{woo98} R.J. Woo {\em et al.}, Phys. Rev. Lett. {\bf 80}
(1998) 456.
\bibitem{milbrath98}
B.D. Milbrath {\em et al.}, Phys. Rev. Lett. {\bf 80} (1998) 452.
\bibitem{kelly} J.J. Kelly, Adv. Nucl. Phys. {\bf 23} (1996) 75.
\bibitem{hummel94} E. Hummel and J.A. Tjon, Phys. Rev. C {\bf 49}
(1994) 21.
\bibitem{ducret94} J.E. Ducret {\em et al.}, Phys. Rev. C {\bf 49}
(1994) 1783.
\bibitem{jan89} 
J. Ryckebusch, K. Heyde, D. Van Neck and M. Waroquier,
Nucl. Phys. {\bf A503} (1989) 694.
\end{thebibliography}
\end{document}